%% file: ms.tex
\documentclass[preprint,3p,times,longtitle, twocolumn]{elsarticle}
\usepackage{mathtools}
\usepackage{textcomp}
\usepackage{amssymb}

\usepackage{siunitx}
\usepackage{xspace}
\usepackage{hyperref}
\usepackage{multirow}
\usepackage{bm}

\usepackage{lineno}

\usepackage{subcaption}
\usepackage{color,soul}

\soulregister\cite7
\soulregister\ref7
\soulregister\pageref7

\journal{Astroparticle Physics}

\begin{document}

\newcommand{\lese}{LESE\xspace} %Low-Energy Starting Events (LESE)
\newcommand{\steve}{STeVE\xspace} %Starting TeV Events (STeVE)

\begin{frontmatter}

\title{Neutrinos below 100 TeV from the southern sky employing refined veto techniques to IceCube data}

\author[christchurch]{M. G. Aartsen\fnref{}}
\author[zeuthen]{M. Ackermann\fnref{}}
\author[christchurch]{J. Adams\fnref{}}
\author[brusselslibre]{J. A. Aguilar\fnref{}}
\author[copenhagen]{M. Ahlers\fnref{}}
\author[stockholmokc]{M. Ahrens\fnref{}}
\author[geneva]{C. Alispach\fnref{}}
\author[erlangen,zeuthen]{D. Altmann\fnref{}}
\author[marquette]{K. Andeen\fnref{}}
\author[pennphys]{T. Anderson\fnref{}}
% ...
\address[brusselslibre]{Universit\'e Libre de Bruxelles, Science Faculty CP230, B-1050 Brussels, Belgium}
% ...
\address[christchurch]{Dept. of Physics and Astronomy, University of Canterbury, Private Bag 4800, Christchurch, New Zealand}
% ...
\address[copenhagen]{Niels Bohr Institute, University of Copenhagen, DK-2100 Copenhagen, Denmark}
% ...
\address[erlangen]{Erlangen Centre for Astroparticle Physics, Friedrich-Alexander-Universit\"at Erlangen-N\"urnberg, D-91058 Erlangen, Germany}
% ...
\address[geneva]{D\'epartement de physique nucl\'eaire et corpusculaire, Universit\'e de Gen\`eve, CH-1211 Gen\`eve, Switzerland}
% ...
\address[marquette]{Department of Physics, Marquette University, Milwaukee, WI, 53201, USA}
% ...
\address[stockholmokc]{Oskar Klein Centre and Dept. of Physics, Stockholm University, SE-10691 Stockholm, Sweden}
% ...
\address[pennphys]{Dept. of Physics, Pennsylvania State University, University Park, PA 16802, USA}
% ...
\address[zeuthen]{DESY, D-15738 Zeuthen, Germany}

\fntext[tokyofn]{Earthquake Research Institute, University of Tokyo, Bunkyo, Tokyo 113-0032, Japan}
\cortext[cor1]{E-mail: analysis@icecube.wisc.edu}

\begin{abstract}
Many Galactic sources of gamma rays, such as supernova remnants, are expected to produce neutrinos with a typical energy cutoff well below 100\,TeV. 
For the IceCube Neutrino Observatory located at the South Pole, the southern sky, containing the inner part of the Galactic plane and the Galactic Center, is a particularly challenging region at these energies, because of the large background of atmospheric muons.
In this paper, we present recent advancements in data selection strategies for track-like muon neutrino events with energies below 100\,TeV from the southern sky. The strategies utilize the outer detector regions as veto and features of the signal pattern to reduce the background of atmospheric muons to a level which, for the first time, allows IceCube searching for point-like sources of neutrinos in the southern sky at energies between 100\,GeV and several TeV in the muon neutrino charged current channel. 
No significant clustering of neutrinos above background expectation was observed in four years of data recorded with the completed IceCube detector. Upper limits on the neutrino flux for a number of spectral hypotheses are reported for a list of astrophysical objects in the southern hemisphere.
\end{abstract}

\begin{keyword}
neutrinos \sep point sources \sep veto techniques

\end{keyword}

\end{frontmatter}

%\linenumbers

%main text

\input{1-Introduction}

\input{2-IceCube}

\input{3-FSS}

\input{4-Strategies}
\input{4.1-STeVE}
\input{4.2-LESE}
\input{4.3-Comparison}

\input{5-Analysis}

\input{6-Summary}

\section*{Acknowledgements}
The IceCube collaboration acknowledges the significant contributions to this manuscript from Rickard Str\"om, David Altmann, and Alexander Kappes. The authors gratefully acknowledge the support from the following agencies and institutions: 
USA -- U.S. National Science Foundation-Office of Polar Programs,
U.S. National Science Foundation-Physics Division,
Wisconsin Alumni Research Foundation,
Center for High Throughput Computing (CHTC) at the University of Wisconsin-Madison,
Open Science Grid (OSG),
Extreme Science and Engineering Discovery Environment (XSEDE),
U.S. Department of Energy-National Energy Research Scientific Computing Center,
Particle astrophysics research computing center at the University of Maryland,
Institute for Cyber-Enabled Research at Michigan State University,
and Astroparticle physics computational facility at Marquette University;
Belgium -- Funds for Scientific Research (FRS-FNRS and FWO),
FWO Odysseus and Big Science programmes,
and Belgian Federal Science Policy Office (Belspo);
Germany -- Bundesministerium f\"ur Bildung und Forschung (BMBF),
Deutsche Forschungsgemeinschaft (DFG),
Helmholtz Alliance for Astroparticle Physics (HAP),
Initiative and Networking Fund of the Helmholtz Association,
Deutsches Elektronen Synchrotron (DESY),
and High Performance Computing cluster of the RWTH Aachen;
Sweden -- Swedish Research Council,
Swedish Polar Research Secretariat,
Swedish National Infrastructure for Computing (SNIC),
and Knut and Alice Wallenberg Foundation;
Australia -- Australian Research Council;
Canada -- Natural Sciences and Engineering Research Council of Canada,
Calcul Qu\'ebec, Compute Ontario, Canada Foundation for Innovation, WestGrid, and Compute Canada;
Denmark -- Villum Fonden, Danish National Research Foundation (DNRF), Carlsberg Foundation;
New Zealand -- Marsden Fund;
Japan -- Japan Society for Promotion of Science (JSPS)
and Institute for Global Prominent Research (IGPR) of Chiba University;
Korea -- National Research Foundation of Korea (NRF);
Switzerland -- Swiss National Science Foundation (SNSF).


\bibliographystyle{elsarticle-num}

\bibliography{References}

\appendix
\clearpage
\input{Appendix}

\end{document}

%% file: 1-Introduction.tex
%% ======================= INTRODUCTION =======================
\section{Introduction} \label{sec:intro}
Cosmic rays below the knee are often thought to be of Galactic origin~\cite{Gaisser:2013bla}. This theory is strengthened by observations by H.E.S.S.~\cite{hess_gc_pevatron} and Fermi~\cite{Ackermann2013} of gamma rays associated with Galactic sources which are mainly found in the southern sky, containing the Galactic center and the majority of the Galactic plane. For many gamma-ray observations, however, it is unclear whether they are produced by interactions of cosmic ray nuclei or leptonic processes. The observation of neutrinos from a Galactic object would be an unambiguous indication of cosmic ray production or interaction in its vicinity. A diffuse astrophysical flux of high-energy neutrinos was first detected with the IceCube detector in 2013~\cite{Aartsen:2013jdh} and recently, for the very first time, convincing evidence for the association of high-energy neutrinos with an astrophysical source (an active galactic nucleus) was presented \cite{IceCube:2018dnn,IceCube:2018cha}. Independent of these successes, observations in particular in the southern sky remain challenging for IceCube because of the large background, on the order of 100 billion muons per year, produced in the atmosphere by cosmic ray interactions.

Neutrino interactions result in two basic patterns in the IceCube detector: tracks and cascades. Muon neutrinos undergoing a charged-current interaction generate a muon which, while traveling through the detector, leaves a track-like pattern of light in the detector. This allows the reconstruction of the direction of the original neutrino with a precision of half a degree to a few degrees in the energy range below \SI{100}{\tera\electronvolt} depending on the track length and energy~\cite{Aartsen:2016oji}. All other neutrino interactions in this energy range lead to a point-like energy deposition generating a spherical (cascade-like) light pattern with a considerably worse angular resolution of up to several tens of degrees~\cite{mesc2017}.

So far, IceCube has used two general strategies to search for emission from point-like sources in the southern sky. In one approach, a cut on the deposited energy retains only events above $\sim$\SI{1}{\peta\electronvolt} (see for example \cite{prl:103:221102}) where the flux of atmospheric muons is substantially suppressed due to their steeply falling energy spectrum. This is challenging for Galactic sources, however, since these are predicted to emit neutrinos only up to a few tens of \SI{}{\tera\electronvolt}~\cite{pr:d78:063004,Vissani:2006tf,Kistler:2006hp}. For the detection of neutrinos with energies below \SI{1}{\peta\electronvolt} from the southern sky, a different approach has been used. By focusing on muon neutrino interactions inside the detector volume and using the outer region as a veto, interactions of astrophysical neutrinos can be distinguished from those of muons and neutrinos of atmospheric origin. This analysis technique was first applied in the discovery of the diffuse astrophysical neutrino flux~\cite{Aartsen:2013jdh}. A modified version of this selection, the Medium-Energy Starting Event (MESE) analysis~\cite{MESE_paper}, has been used to improve the sensitivity of the IceCube detector to point-like neutrino sources in the southern sky for energies below \SI{100}{\tera\electronvolt}, with most of the gain applying to neutrinos above \SI{50}{\tera\electronvolt}. In addition, a search for neutrino sources in the southern hemisphere was performed looking for cascade-like event signatures~\cite{mesc2017}, resulting in an improved sensitivity to spatially extended sources and sources that follow a soft energy spectrum.

To improve the sensitivity of track-like searches in the range from  \SI{100}{\tera\electronvolt} down to \SI{100}{\giga\electronvolt}, two new selection strategies, \steve (Starting TeV Events) and \lese (Low-Energy Starting Events), have been developed and are presented in this paper. These strategies use several veto techniques with the aim of keeping the fiducial volume of the detector large while reducing the background from atmospheric muons.

The paper is structured as follows: In Section \ref{sec:icecube} we introduce the IceCube Neutrino Observatory, including a brief discussion of the conditions used to form event triggers relevant for veto-based selection strategies. Section \ref{sec:fss} describes the Full Sky Starting (FSS) online event filter developed for online\footnote{Processing in the IceCube Laboratory at the South Pole.} selection of low-energy events from the southern sky, and in Section \ref{sec:selections} the \steve and \lese event selections are presented (in-depth discussions can be found in \cite{Altmann:thesis2017} and \cite{strom:thesis2015}, respectively). The performance of these event selections is demonstrated in Section \ref{sec:psanalysis}, where they are applied to a search for point-like sources in the southern sky using four years of data from the completed IceCube detector. This section also includes a discussion of the major systematic uncertainties of the study. We conclude with a summary and an outlook in Section \ref{sec:summary}.

%% file: 2-IceCube.tex
\section{The IceCube Neutrino Observatory} \label{sec:icecube}

\hyphenation{para-boloid}

\begin{figure}[ht!]
  \centering
  \includegraphics[width=220pt]{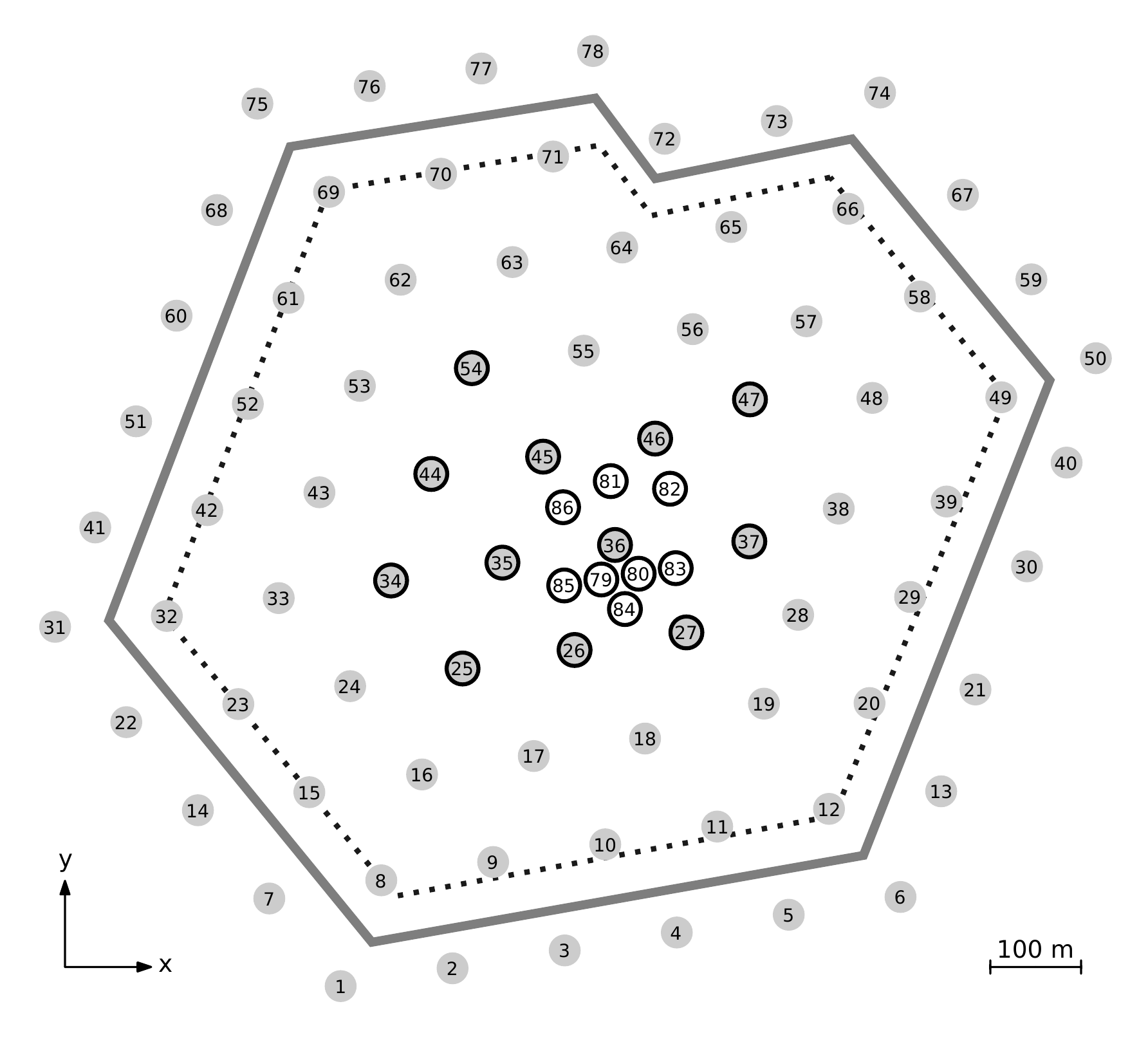}
  \caption{Top view of the IceCube detector with dots representing the positions of the numbered sensor strings. While gray dots indicate strings with nominal string spacing, white dots with a black border indicate strings with a higher DOM density, forming a low-energy extension of IceCube called DeepCore. The strings marked with a black border are used to form one of the primary IceCube triggers as explained in the main text. The polygons define different volumes which are used in the FSS filter (solid grey) and the \steve and \lese event selections (dashed grey) described in Section \ref{sec:fss} and \ref{sec:selections}, respectively.} \label{fig:strings_veto}
\end{figure}

The IceCube Neutrino Observatory~\cite{IceCube:2017} is a multi-purpose observatory with a broad range of scientific topics including, as one of its main goals, the understanding of the origin of cosmic rays and their acceleration mechanisms~\cite{Halzen:2010yj}. The main instrument is a cubic-kilometer neutrino detector consisting of 5160 optical sensors embedded in the ultra-clear Antarctic glacial ice at the geographic South Pole at depths between 1450\,m and 2450\,m. It consists of 86 cables, called strings, each housing 60 optical sensors, referred to as Digital Optical Modules (DOMs)~\cite{Abbasi:2008aa}. Seventy-eight of the strings form a triangular grid with a nominal string spacing of \SI{125}{\metre}~\cite{Abbasi:2010vc} and a vertical DOM-to-DOM spacing of \SI{17}{\metre}. The center of the regular grid is augmented by an additional 8 strings, forming a low-energy extension called DeepCore with a higher DOM density~\cite{IceCube:2011ym}. Fig.~\ref{fig:strings_veto} shows an illustration of IceCube as seen from above, where the position of the strings of the regular grid is marked with gray dots while DeepCore strings are indicated with white dots with black contours.
IceCube observes neutrinos by detecting the Cherenkov light emitted by charged particles produced in the interaction of neutrinos with nuclei in the ice or the nearby bedrock. The Cherenkov light is detected by a 10-inch photomultiplier tube (PMT) housed inside each DOM. If two neighboring or next-to-neighboring DOMs in a string cross the discriminator threshold (representing 25\% of the average amplitude of one photoelectron) within a $1\,\mu$s time window, the signals qualify as a Hard Local Coincidence (HLC). DOMs that do not qualify for HLC, but that cross the discriminator threshold, are defined as having a Soft Local Coincidence (SLC); these are particularly useful for the improvement of the reconstruction and veto efficiency of low-energy events. The majority of the SLCs are due to pure noise, which is why dedicated noise cleaning algorithms are applied for both the angular and energy reconstructions (see Section \ref{ssec:noisecleaning}).

The digitized signals for all DOMs with hits above the discriminator threshold are sent to the surface, including the full PMT waveform information in case of HLC DOMs. For SLC DOMs, reduced waveform information is sent (for details see \cite{IceCube:2017}). The number of DOMs reporting HLC is used to trigger a readout of the entire IceCube detector: three different triggers, called SMT-8, SMT-3, and StringTrigger, are used in the studies presented in this paper. The leading trigger condition, SMT-8, is formed when at least 8 DOMs are in HLC within a sliding time window of $5\,\mathrm{\mu}$s. Similarly, SMT-3, only active in the center region of the detector, indicated by the 20 strings shown with a black border in Fig.~\ref{fig:strings_veto}, is formed when at least 3 DOMs are in HLC within a sliding time window of $2.5\,\mathrm{\mu}$s~\cite{IceCube:2017}. The StringTrigger aims at catching almost vertical events and is formed when 5 out of 7 adjacent DOMs on the same string are in HLC within a sliding time window of $1.5\,\mathrm{\mu}$s. Overlapping triggers are merged and the resulting trigger window is padded by -4\,\SI{}{\micro\second} and +6\,\SI{}{\micro\second}, forming a single event. The triggered events are dominantly ($>99.99\%$) atmospheric muons, with the contribution from atmospheric neutrinos approximately a factor $10^{6}$ smaller. 

Detailed Monte Carlo simulations are used to evaluate the response of the detector to neutrinos and atmospheric backgrounds, and to simulate the optical noise in the modules (for more information about the details of the Monte Carlo simulation used, see \cite{Aartsen:2016xlq, Heckcorsika:a, Koehne20132070} and references therein).

\subsection{Noise cleaning} \label{ssec:noisecleaning}

Dedicated cleaning algorithms are used to identify and remove hits related to noise in the detector. Angular reconstructions are quite sensitive to noise hits, and thus require a strict cleaning. In contrast, semi-isolated hits of low quality contain important information for use in a veto. This is why the reconstructions of neutrino interaction vertices are performed using a less strict cleaning. The cleaning and subsequent reconstructions are based on the collection of observed pulses in each event, where each pulse is defined by the charge and leading edge time as extracted from the corresponding waveforms using an iterative unfolding algorithm with pre-defined templates~\cite{icecube_ereco}.

For reconstruction of the neutrino interaction vertex, pulses are first cleaned using a time-window cleaning algorithm in which pulses are rejected if they are outside the range [-4, 6]\,\SI{}{\micro\second} relative to the earliest recorded trigger in the event. In a second step, both timing and spatial information are used to remove pulses that are isolated from the main clusters of hits. This cleaning retains 96\% of the physics hits and about 18\% of the noise hits.

For the angular reconstructions, an iterative causality cleaning algorithm, initially considering only a clean subset of HLC pulses, is applied. Starting from this core, pulses are added if they are within a specified time and radius of the seed pulses. This is followed by the application of a time-window cleaning algorithm rejecting pulses outside the range [-4, 10]\,\SI{}{\micro\second}, relative to the earliest trigger of the event. This cleaning is slightly stricter than the cleaning used for the reconstruction of the interaction vertex, and keeps 92\% of the physics hits, while rejecting 97\% of the noise hits.

\subsection{Reconstruction techniques} \label{ssec:reco}

The timing and location of the DOMs participating in an event provide the most important information in the determination of the most likely direction and position of the muon in the detector, which is in turn the best available proxy for the neutrino arrival direction; the neutrino-induced muons are highly boosted in the forward direction but are produced with a small angle, approximately $\langle\psi_{\nu\mu}\rangle\approx0.7^\circ/(E_\nu/\,\mathrm{TeV})^{0.7}$~\cite{Learned:2000sw}, compared to the primary neutrino direction. 
A simple algorithm~\cite{Aartsen:2013bfa} is used as a first guess, while more advanced algorithms~\cite{reconstruction} are applied following several stages of cuts, when the number of events has been reduced significantly. The advanced algorithms use the expected photon arrival time distribution from the track hypothesis, taking into account effects of the ice~\cite{Aartsen:2013rt}.

Early levels of event selections use the result from an iterative fit based on a single photoelectron (SPE) PDF, approximating the timing distribution of the Cherenkov photons arriving at a given PMT using only the time of the first recorded pulse for each participating DOM. The first recorded pulse is likely to be the least scattered and hence contains the most information about the true track direction. At higher levels of the selections, the so-called multi-photoelectron (MPE) likelihood is used which, in addition to the time and charge of the first photon, uses the sum of charges of all subsequent photons. Furthermore, while the SPE-based reconstruction uses an analytical approximation of the timing distribution of the Cherenkov photons arriving at a given PMT, in the MPE algorithm a parametrization of a Monte Carlo simulation of the photon transport in ice is used~\cite{Whitehorn:2013nh}. The resulting multi-dimensional spline tables are used together with a depth-dependent model of the optical properties of the ice~\cite{Aartsen:2013rt}. The SPE-based reconstruction is used in the online event selection described in Section \ref{sec:fss}, while the MPE-based reconstruction is used in a likelihood analysis to search for clustering of signal-like events.

The angular uncertainty of each event is estimated by fitting a paraboloid to the likelihood space around the reconstructed direction obtained  from the MPE-based algorithm~\cite{Neunhoffer:2004ha}. Since the paraboloid algorithm often underestimates the true angular uncertainty, an energy-dependent correction, determined from simulation, is applied. In early steps of the \lese selection, an additional cut is applied on the angular uncertainty, estimated using the Fisher information matrix of the provided track reconstruction.

The reconstructed vertex position of the neutrino interaction is a key parameter in the low-energy veto-based event selections described in Section \ref{sec:selections}. A simple \emph{first-guess} approach is used in the initial event filter: the point of earliest photon emission is estimated by projecting all hits within 200 m of the input seed track onto that track along the Cherenkov angle~\cite{Hulss:2010zz}.

At higher levels of the selections, a more advanced vertex reconstruction algorithm is used, that estimates both the starting and stopping point of the muon. The algorithm minimizes a likelihood considering the probability to not observe photons, given an infinite/finite track hypothesis. Further details of this procedure are presented in~\cite{Hulss:2010zz}. Additional variables are derived from these reconstructions, for example the reduced log-likelihood value indicating the overall quality of the fit, and the ratio between the individual likelihoods for the finite and infinite track hypotheses. These variables are used in the \lese selection, see Section \ref{ssec:lese}.

%% file: 3-FSS.tex
\section{Full Sky Starting (FSS) online filter} \label{sec:fss}

The FSS online filter is configured to select track-like neutrino-candidate events that interact inside the IceCube detector. The filter uses the outermost DOMs of IceCube as a veto. The main background consists of atmospheric muons, entering the detector from the outside. The filter targets events with energies below 100 TeV, but does not include an explicit cut on any energy-related variables. However, high-energy background events are more likely to leave traces in the veto region, due to the increased Cherenkov emission at higher energies, and hence be rejected.

\begin{figure}[ht!]
        \centering
        \begin{subfigure}{210pt}
	    \includegraphics[width=220pt]{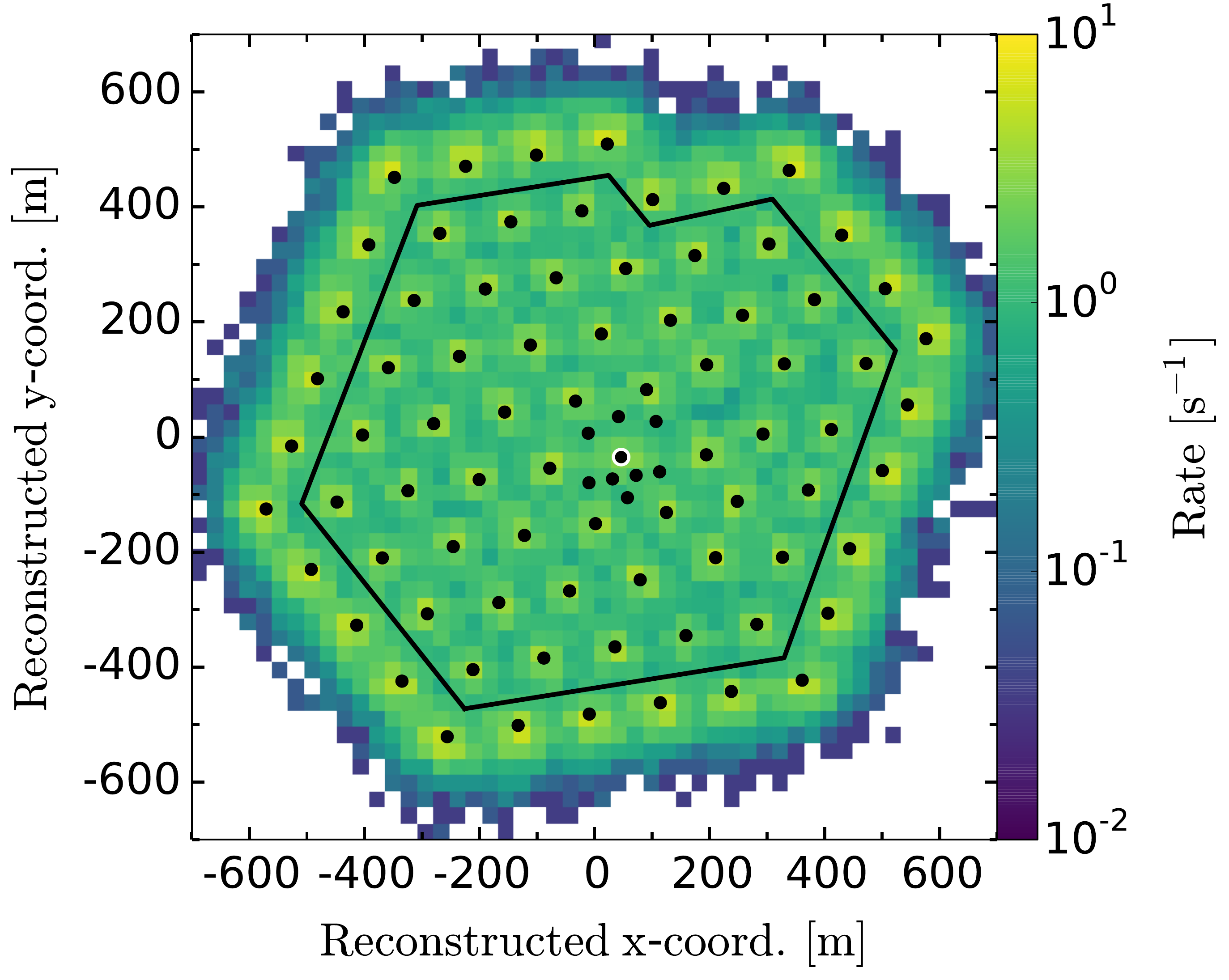}
	    \caption{Top view of the IceCube detector where string positions are shown as black dots. The polygon shape represents the containment criterion for the reconstructed interaction vertices (see also Fig. \ref{fig:strings_veto}).}
        \end{subfigure}\\
        \vspace{4mm}
        \begin{subfigure}{210pt}
	    \includegraphics[width=220pt]{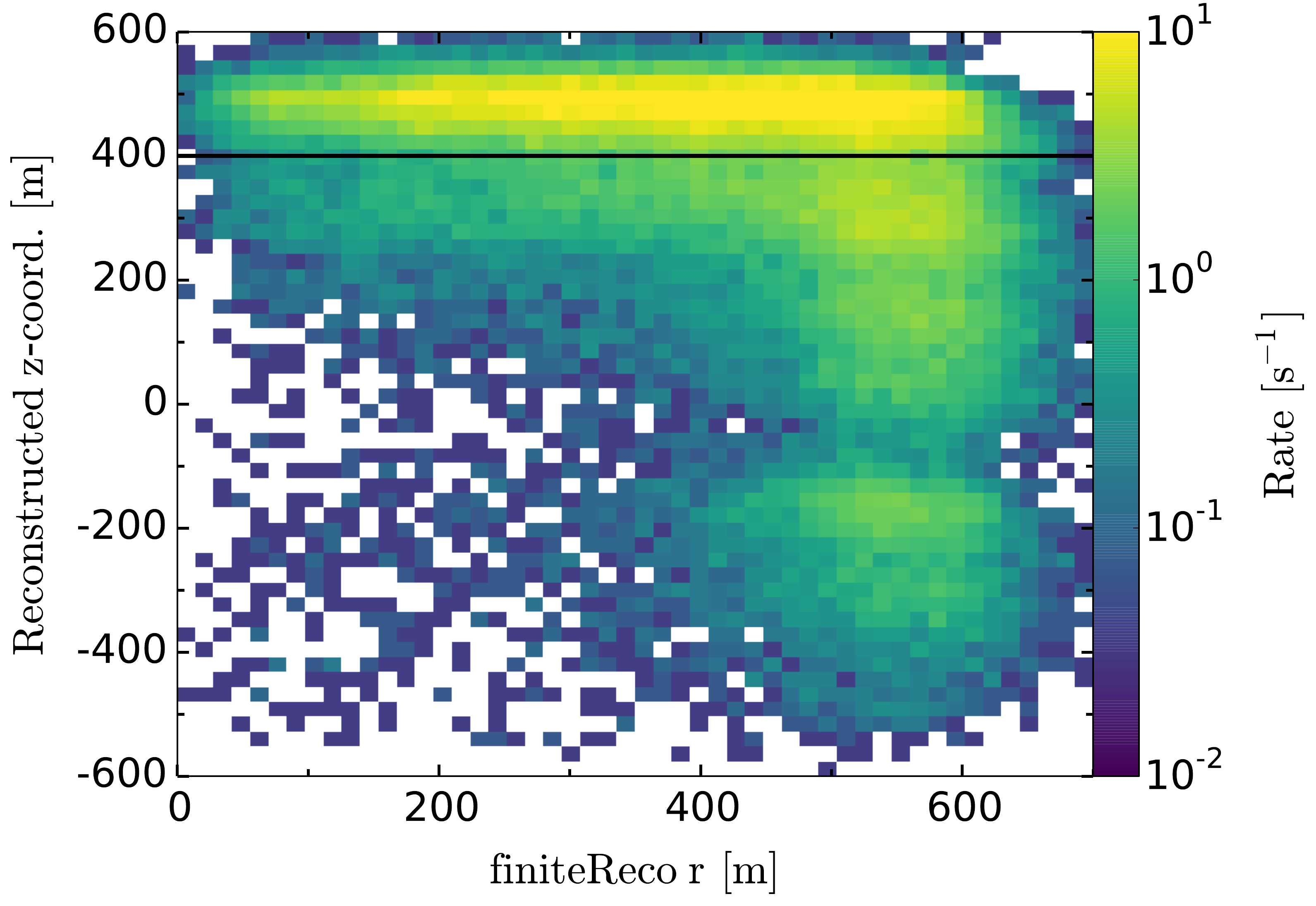}
	    \caption{Side view of the detector where the horizontal black line indicates the containment criterion used in the top veto of the FSS filter. The horizontal axis (``reconstructed r'') shows the position of the vertex in the xy-plane, i.e. $\sqrt{x^2+y^2}$. Note that z = 0 corresponds to the center of the detector.}
        \end{subfigure}
        \caption{Distributions of reconstructed interaction vertices for experimental data before the application of the FSS filter. The color (z-axis) indicates the event rate.}
        \label{fig:fss}
\end{figure}

The FSS filter operates in a two-step approach. The first step is a veto based on the hit pattern of DOMs fulfilling the HLC condition described in Section \ref{sec:icecube} as a means for fast separation of starting and non-starting events: no DOMs satisfying HLC are allowed in the five top-most DOMs in each of the 78 strings of the large triangular grid. Furthermore, the first HLC is not allowed on any of the strings of the outer-most\footnote{The outer-most layer of IceCube includes the following strings: 1-7, 13, 14, 21, 22, 30, 31, 40, 41, 50, 51, 59, 60, 67, 68, 72-78 (see Fig. \ref{fig:strings_veto} for reference).} layer as seen from the top. The second step of the filter uses the reconstructed starting vertex from the simple \emph{first-guess} algorithm described in Section \ref{ssec:reco}. An event with a reconstructed vertex outside the solid polygon in Fig.~\ref{fig:strings_veto} and/or in the top \SI{100}{\metre} of the detector volume is discarded. Fig.~\ref{fig:fss} shows the distribution of reconstructed vertices for experimental data before application of the FSS filter.
The horizontal bands visible in the side view indicate regions of lower sensitivity to optical light. This is caused by dust layers in the ice with an increased absorption and scattering of photons~\cite{Ackermann:2006pva}.
The passing rate of the FSS filter is about 190\,$\mathrm{s}^{-1}$, whereas the global IceCube trigger rate varies from 2.5$\cdot10^3\,\mathrm{s}^{-1}$ to 2.9$\cdot10^3\,\mathrm{s}^{-1}$~\cite{IceCube:2017}.

Simpler starting event filters were used during the construction phase of IceCube with partial detector configurations. These targeted specific regions of interest, such as the Galactic center (as used in \cite{Aartsen:2015xej}), or focused on low-energy events detected in the denser DeepCore array with a substantially smaller active volume (as used in \cite{Aartsen:2016pfc}). In contrast, the FSS filter uses a large part of the IceCube detector as well as the DeepCore array to accept events from the entire southern sky.

%% file: 4-Strategies.tex
\section{Selection strategies} \label{sec:selections}
Two event selections, \steve and \lese, have been developed to take advantage of the events that pass the FSS filter. While they focus on different energy ranges, both aim at an event sample enriched in neutrino candidate events interacting inside the instrumented detector volume while maintaining an angular resolution on the order of one degree. The latter requirement is particularly important for identifying clustering among neutrinos as well as possible correlations of such with the location of established sources of electromagnetic radiation. The sparse DOM instrumentation poses a major challenge for using vetoes to suppress atmospheric muons with energies below 10 TeV. Atmospheric muons that reach the inner detector volume without leaving a detectable signal in the outer regions become an irreducible background for point-source searches, similar to atmospheric neutrinos.

The idea for \steve is based on the event selection presented in \cite{MESE_paper} (MESE) with the aim of lowering the energy threshold further, below \SI{50}{\tera\electronvolt}. The second strategy, \lese, aims at selecting track-like events with energies as low as 100\,GeV, leveraging the experience gained with veto-based selection techniques in searches for dark matter~\cite{Aartsen:2015xej,Aartsen:2016pfc,Aartsen:2017if}. Common to both strategies is the suppression of events with multiple atmospheric muons. Such coincident events are particularly challenging to veto and reconstruct as a whole as they do not fit the hypothesis of a single muon. Therefore, such events are split into separate single-muon events utilizing the spatial and temporal pattern of hits.

Throughout the event selections we use a sub-sample of experimental data to represent the atmospheric muon background, while simulated muon-neutrino events were used to describe the signal. Simulated atmospheric muons were used to verify the overall shape of each variable compared to the experimental data sample: variables with significant discrepancy beyond the statistical uncertainty of the samples were excluded from the analyses.

%% file: 4.1-STeVE.tex
\subsection{Starting TeV Events (\steve)} \label{ssec:steve}

This event selection strategy exploits the difference in the observed photon pattern of bundles of low-energy muons compared to individual high-energy muons. The selection focuses on identifying starting events from the southern hemisphere at energies between 10\,TeV and 100\,TeV. In order to reduce the event rate to a level where sophisticated reconstructions for the extraction of detailed track parameters can be used, a cut on a first-guess energy estimator is applied. The energy is reconstructed by evaluating the average light yield of a hypothetical infinite muon track with emission of photons at the Cherenkov angle. The expected number of photons for each DOM is compared to the observed photon pattern and the difference is minimized using a likelihood. The ice properties are described analytically, neglecting depth dependency~\cite{icecube_ereco}. While events with a reconstructed energy larger than \SI{10}{\tera\electronvolt} pass, events with a lower reconstructed energy, down to \SI{1}{\tera\electronvolt}, only pass the selection if the maximal distance between a pair of HLC DOMs in the event exceeds \SI{150}{\metre}, or if the first DOM with an HLC pulse is not on the second outermost layer of the detector marked by the dotted line in Fig.~\ref{fig:strings_veto}. Note that events with the first HLC on the outermost layer are already removed by the FSS filter.

\begin{figure}[ht!]
	\centering
	\begin{subfigure}{220pt}
	    \includegraphics[width=220pt]{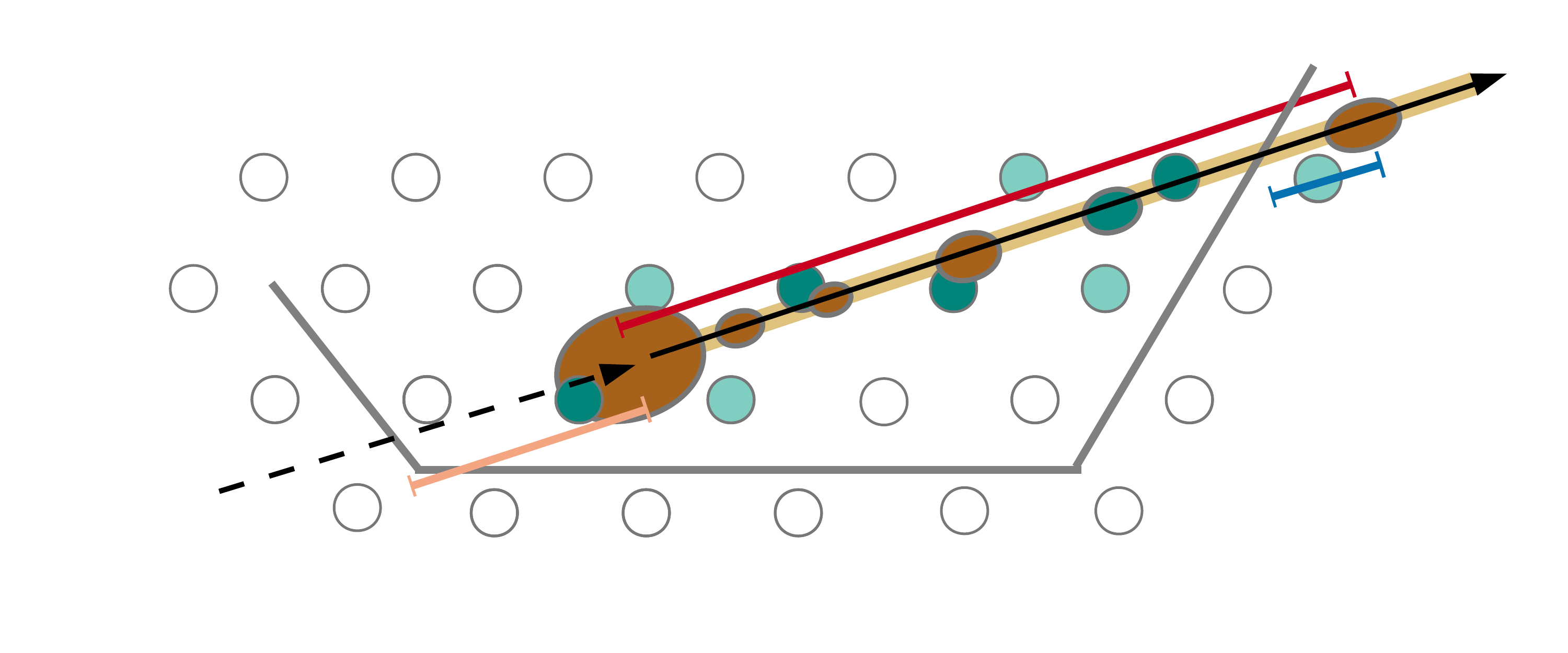}
	    \caption{Starting muon track}
	\end{subfigure}
	\begin{subfigure}{220pt}
	    \includegraphics[width=220pt]{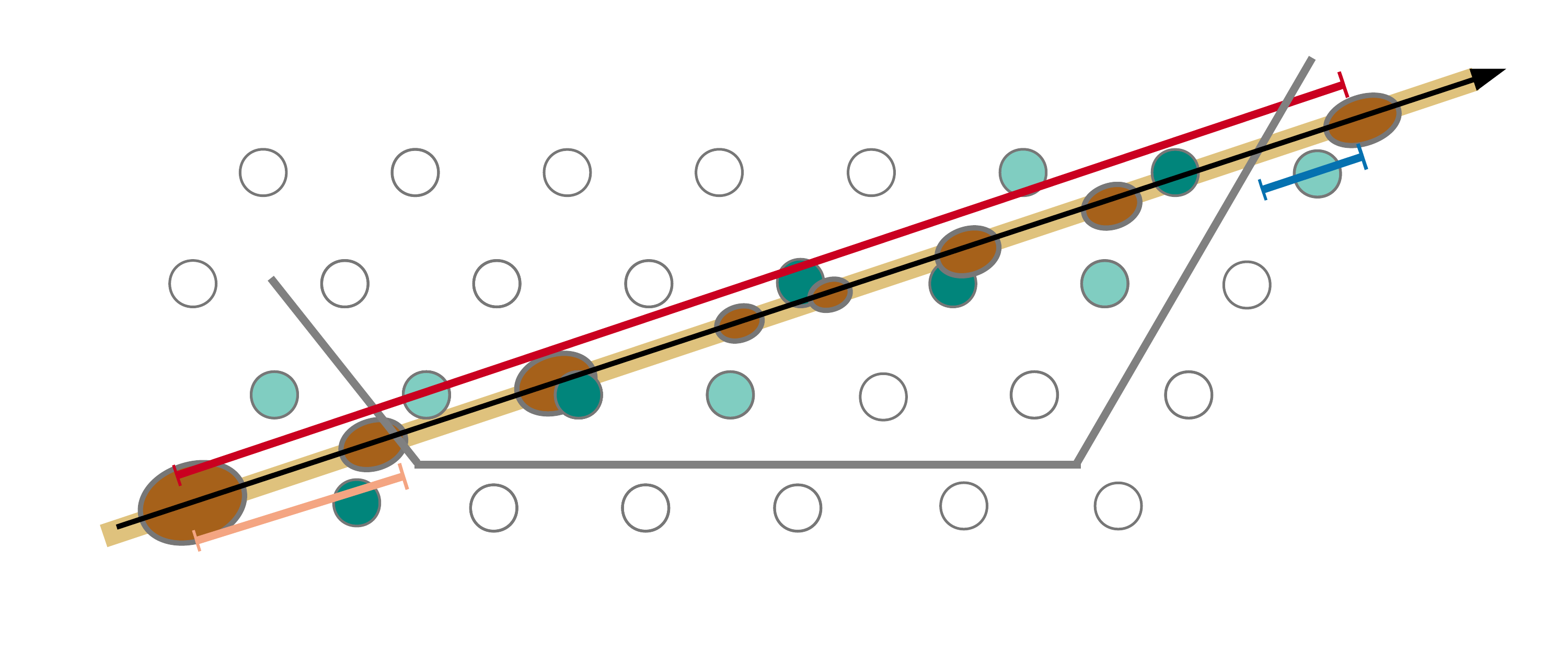}
	    \caption{Throughgoing muon track}
	\end{subfigure}
	        \vspace{3mm}
	\begin{subfigure}{220pt}
	    \includegraphics[width=220pt]{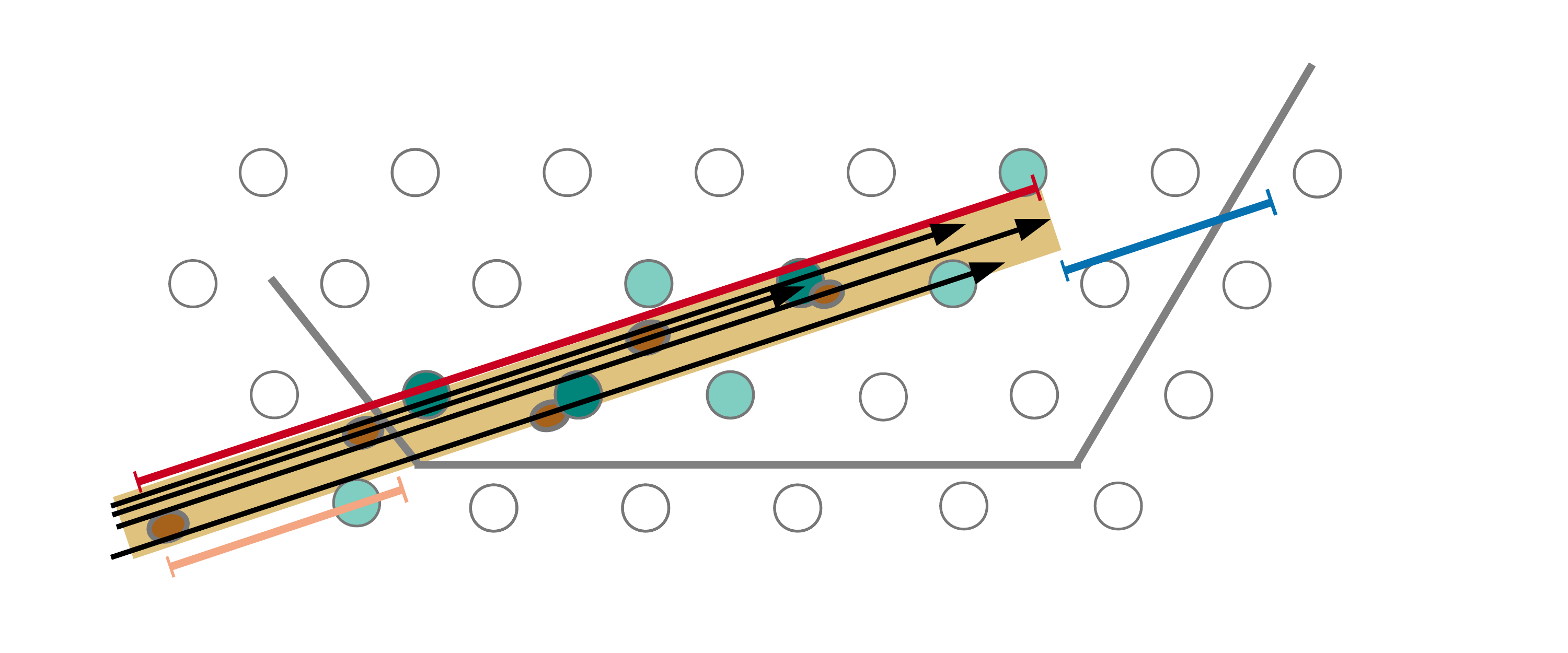}
	    \caption{Low-energy muon bundle}
	\end{subfigure}
	\begin{subfigure}{220pt}
	    \includegraphics[width=220pt]{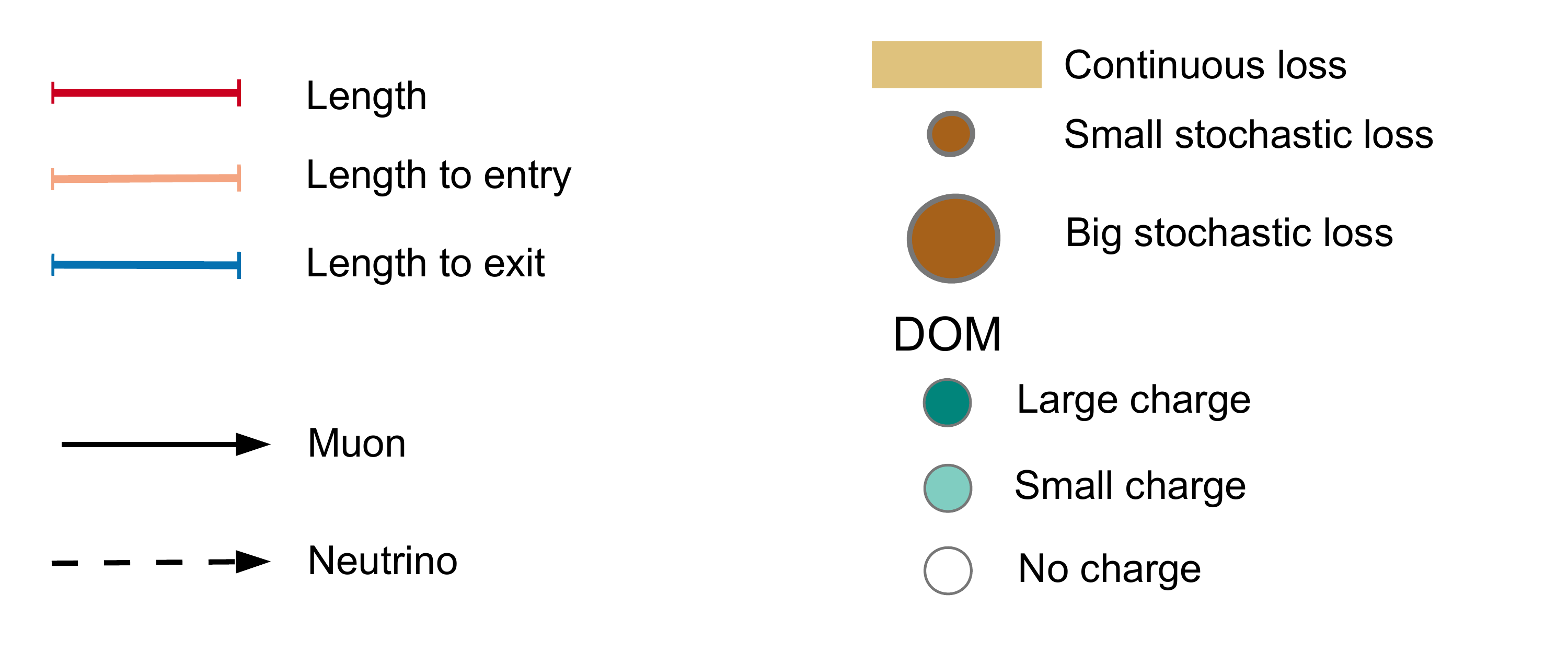}
	\end{subfigure}
\caption{Top view of different track-like events in the IceCube detector, showing quasi-continuous and stochastic energy losses as well as the definition of three useful track length variables. Figure adapted from \cite{Altmann:thesis2017}.}
\label{fig:event_topologies}
\end{figure}

After this initial rejection more sophisticated direction and energy reconstruction algorithms are applied, aiming for efficient separation of neutrino-induced single muon tracks that start inside the detector volume from background muon tracks passing undetected through the veto layers. The latter mainly consists of atmospheric muon bundles with a smooth energy loss distribution. In contrast the energy loss of individual high-energy muons is dominated by stochastic losses, as illustrated in Fig.~\ref{fig:event_topologies}.

The events are reconstructed using the segmented energy-loss algorithm discussed in \cite{icecube_ereco}. This determines the energy deposition of the event by fitting stochastic energy losses in segments along the reconstructed track, in this case with a spacing of \SI{15}{\metre} between each segment. The position and energy of the fitted losses are used to calculate several length and energy parameters of the track (Fig.~\ref{fig:event_topologies}). The reconstructed \emph{length} in the detector is defined by the distance between the first and last energy loss along the track. Two additional track variables are derived with respect to the veto-boundary depicted with a grey solid line (see also Fig.~\ref{fig:fss}): the \emph{length-to-entry}, defined as the distance between the first energy loss along the track and the entry point of the track through the veto-boundary, and the \emph{length-to-exit}, defined as the distance between the last energy loss along the track and the exit point of the track through the veto-boundary. As muons with \si{\tera\electronvolt} energies have average propagation lengths well above the size of the IceCube detector, a muon generated inside the detector will generally leave the detector. On the other hand, an atmospheric muon can lose most of its energy before reaching the detector and therefore stop inside. An energy estimate of the total deposited energy for each event is defined as the sum of the individually reconstructed energy depositions in each segment, excluding depositions reconstructed outside of the instrumented detector volume.

\begin{figure}[ht!]
  \centering
  \includegraphics[width=220pt]{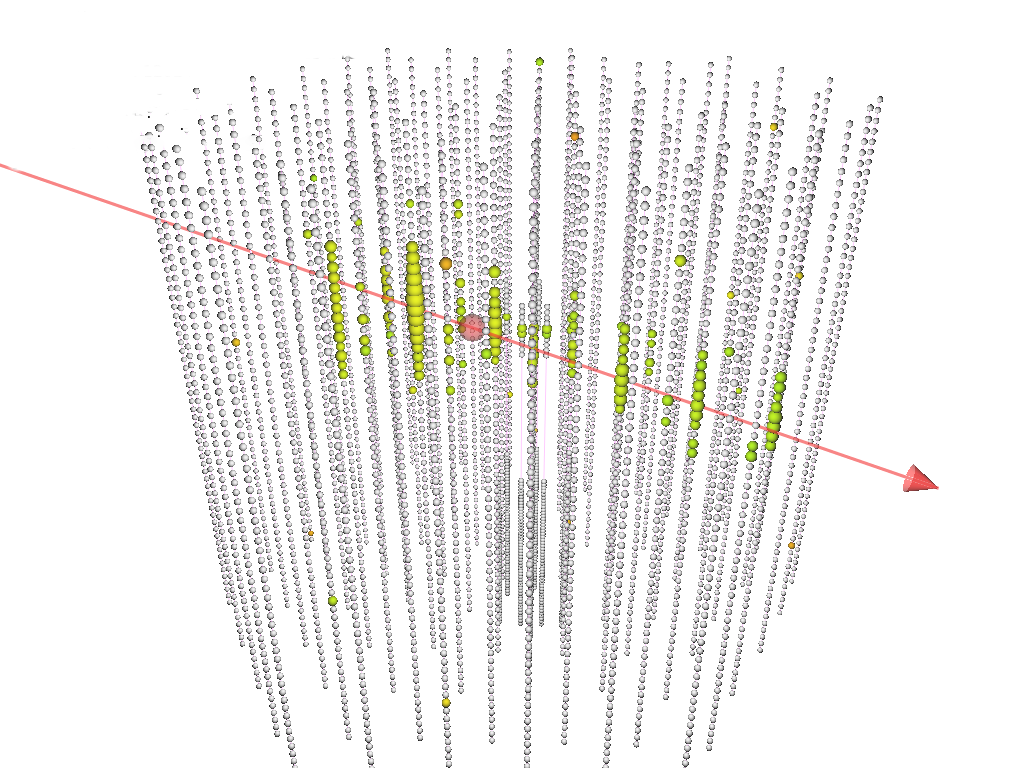}
  \caption{An example starting event candidate of the \steve selection.}
  \label{poster_boy}
\end{figure}

Apart from the quantities discussed above, additional parameters describing the event, for example the charge weighted mean of DOM positions, have been found to provide separation power between signal and background. In total, 19 observables were selected. Details about these observables can be found in~\cite{Altmann:thesis2017}. The observables were used as the inputs to a binary classifier which made use of both boosting and pruning, producing an event score between $-1$ and $1$ indicating whether the events are background- or signal-like, respectively. The classifier was trained using experimental data as background and well reconstructed simulated muon neutrino events with energies from \SIrange{1}{100}{\tera\electronvolt} as signal. The latter interact inside the detector volume indicated by the dotted line in Fig. \ref{fig:strings_veto}. Events below a threshold value in the classifier output are discarded. The threshold is chosen to yield optimal sensitivity for point-like sources with a spectrum described by an $\mathrm{E}^{-2}$ power law and a cutoff around \SI{100}{\tera\electronvolt}. An example event from the final event selection is displayed in Fig. \ref{poster_boy}.

%% file: 4.2-LESE.tex
%% ======================= LESE =======================
\subsection{Low-Energy Starting Events (\lese)} \label{ssec:lese}

This event selection focuses on identifying starting events from the southern hemisphere with energies below 10\,TeV. It consists of several consecutive steps of cuts and data processing, each focusing on a different task. The initial steps deal with the overall data quality, for example by removing events with hits on fewer than three strings, the minimum required to resolve the azimuthal direction. This is followed by the application of several different veto methods as well as a final selection through a machine-learning algorithm. At higher selection levels, with considerably reduced event rates, more advanced and time-consuming reconstructions are used.

The event selection was optimized for signal neutrinos interacting inside the volume defined by the dotted polygon in Fig. \ref{fig:strings_veto} with $z\leq300$ m and a power-law spectrum with an exponential cut-off at 10\,TeV, $\mathrm{E}_\nu^{-2}e^{-\mathrm{E}_\nu/10\ \mathrm{TeV}}$. Up to the application of the machine-learning algorithm, each cut was optimized to maximize the significance approximated as $S/\sqrt{B}$, where $S$ and $B$ represent the number of signal- and background-like events respectively. 

The \lese selection applies a stricter cut on the location of the reconstructed interaction vertex compared with the one used in the FSS filter, removing events with vertices in the top \SI{250}{\metre} of the detector volume~\cite{strom:thesis2015}. Furthermore, the quality of the sample in terms of angular resolution is improved by cutting on the uncertainty associated with the reconstructed track. Several variables connected to the starting vertex and length of the track (strongly correlated to the angular uncertainty) are used to improve both the sample reconstruction quality and the signal/background separation. This includes a cut on the distance between the reconstructed starting vertex and the charge weighted mean of the DOM positions from the latest 25\% of the recorded hits, as well as a cut on the distance between the starting vertex and the horizontal projection of the point on the reconstructed track at the top of IceCube, located \SI{1450}{\metre} below the ice surface.

Furthermore, two additional variables connected to the detected charge pattern are used: the first measures the mean deviation of the z-coordinate of all pulses to the charge weighted mean value of the z-coordinates of the first 25\% of the pulses in time. Here, a starting muon track is expected to have a small value, as the majority of the charge is deposited relatively close to the interaction point. The second variable measures the time until 75\% of the detected charge is collected. As a starting track leaves the detector, very long accumulation times would indicate the presence of one or more non-starting tracks.

\begin{figure}[ht!]
        \centering
        \begin{subfigure}{220pt}
	    \includegraphics[width=220pt]{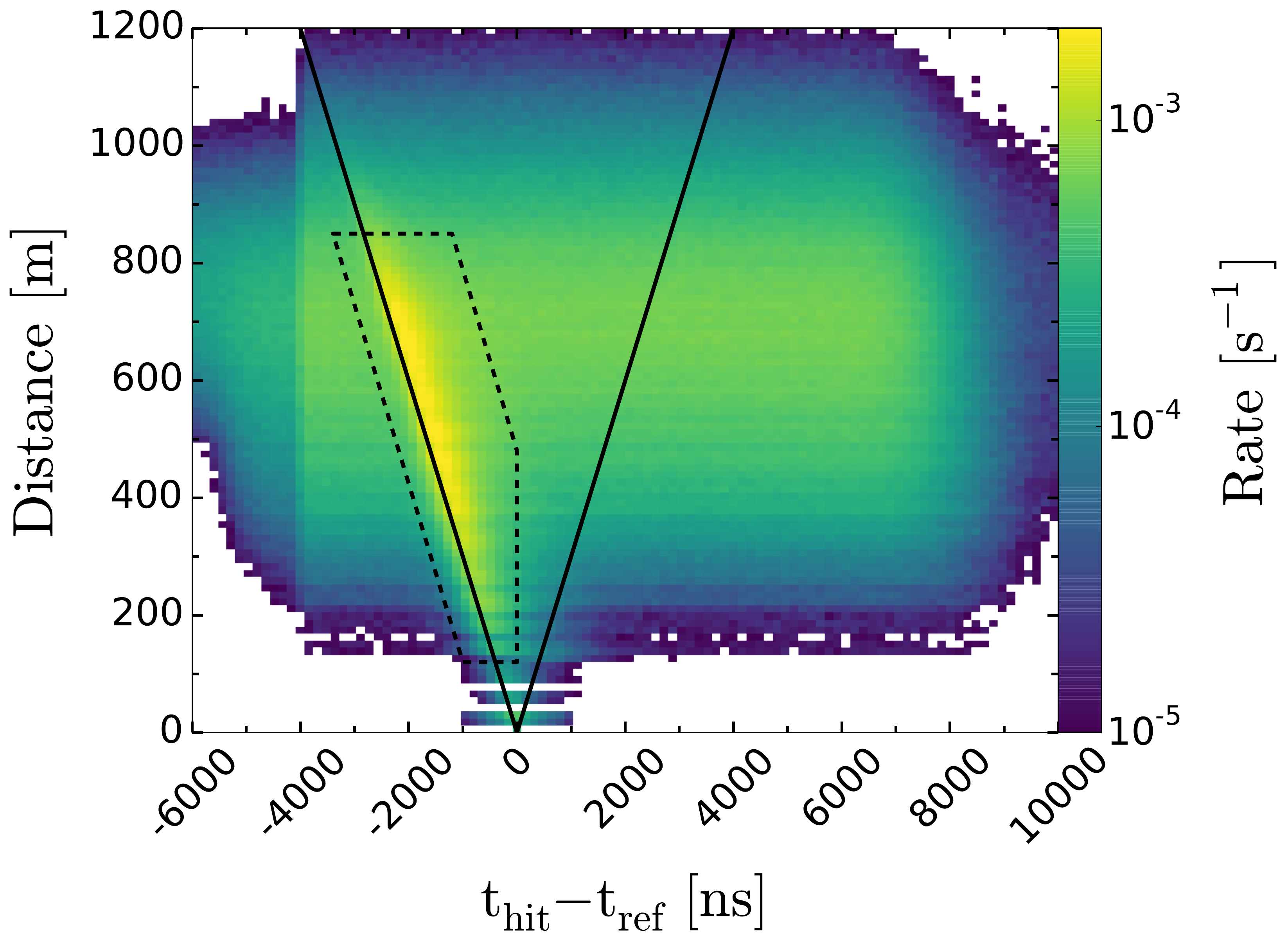}
	    \caption{Experimental data.}
        \end{subfigure}\\
        \vspace{4mm}
        \begin{subfigure}{220pt}
	    \includegraphics[width=220pt]{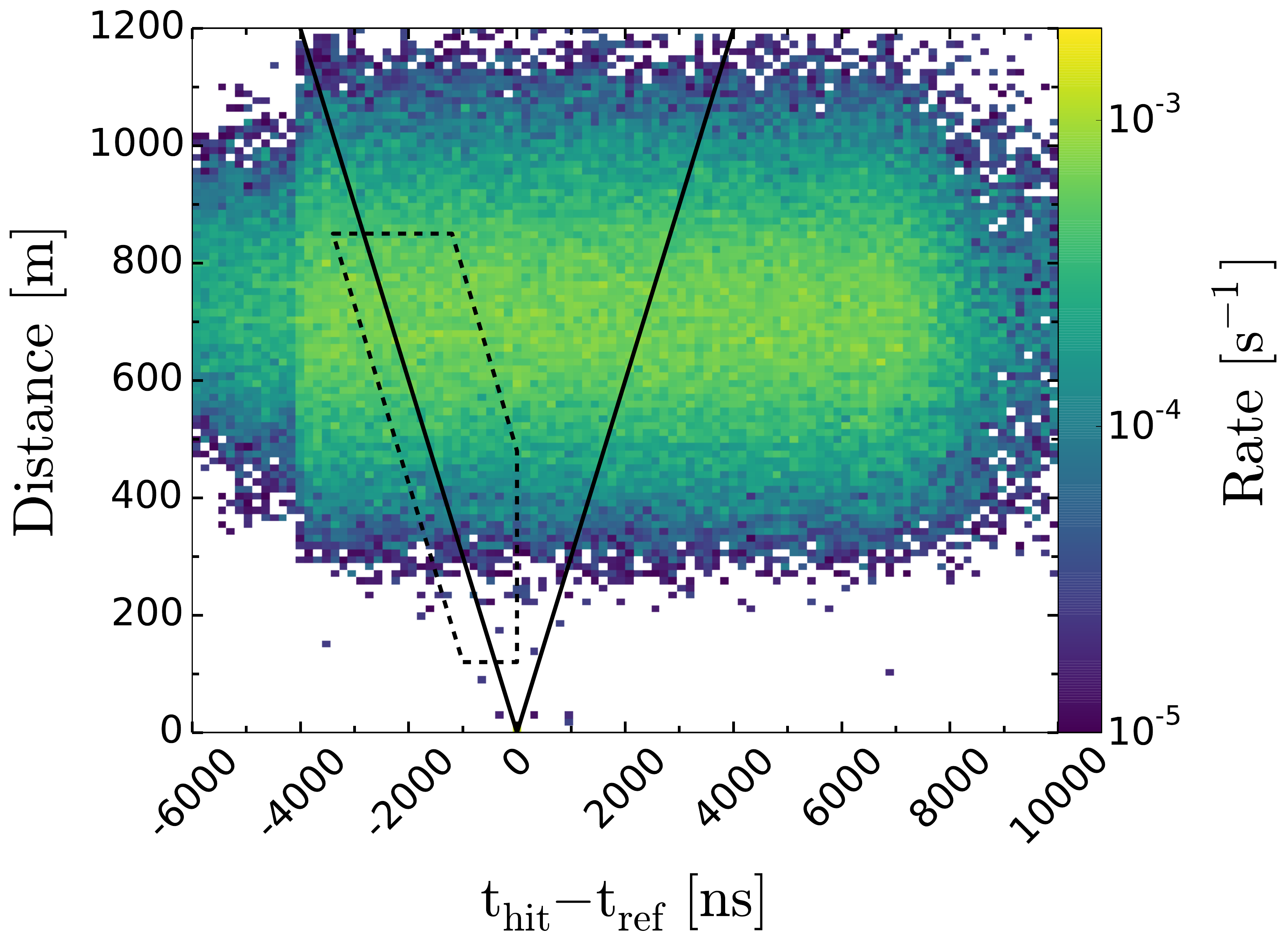}
	    \caption{Simulated, truly down-going, muon neutrinos interacting inside the volume defined by the dotted polygon in Fig. \ref{fig:strings_veto} and $z\leq300$ m.}
        \end{subfigure}
       \caption{Distance in space and time between SLC hits in the top veto region and the first HLC hit in the fiducial volume. The two solid black lines converging at the position of the reference hit (0,0), illustrate the light-cone. The color (z-axis) indicates the event rate.}
        \label{fig:crveto_top}
\end{figure}

Next, the selection focuses on using techniques to improve the overall veto efficiency. At low energies, down-going atmospheric muons can pass the veto layer without depositing sufficient energy to fulfil the HLC condition, but will instead leave traces of SLC hits. The information of these pulses can be used to reject background-like events and is exploited in the subsequent steps of the selection.

%LLH Veto
In the first method, all pulses within a radius of \SI{350}{\metre} around the seed track and with a location on the incoming side of the reconstructed interaction vertex are considered. The compatibility of these pulses with an infinite track hypothesis is evaluated using a maximum-likelihood algorithm. A high likelihood value indicates that some pulses are connected to the track, which is why these events are rejected.

%CRVeto
The second method relies solely on the detected hit pattern with no explicit dependence on a particular track hypothesis. Two veto regions are defined in the detector volume: a top veto and a side veto. The top veto includes the 12 top-most DOMs of each of the 78 non-DeepCore strings. For the side veto, the two outermost layers of strings are used. Either one of these vetoes is applied, depending on the location of the event in the detector.

The separation in space, $\delta_r$ ('Distance' in Fig. \ref{fig:crveto_top}), and time, $\delta_t=\mathrm{t}_\mathrm{hit}-\mathrm{t}_\mathrm{ref}$, is calculated between each SLC hit in the veto and the first HLC hit in the fiducial volume. The latter is defined as the volume of IceCube with DOMs that are not participating in the veto region. If a causal connection exists, the hits are expected to line up approximately along the light-cone describing a particle traveling at the speed of light in vacuum, $c$. Fig. \ref{fig:crveto_top} displays the distribution of potential veto hits relative to the reference hit for the top veto. Note that the sign of $\delta_t$ is defined such that negative values identify hits occurring before the reference time. A clear correlation along the light-cone is seen for experimental data, indicating events leaking in through the initial filter. Note that the edge at \SI{-4}{\micro\second} is caused by the different time windows used for the active detector triggers. Furthermore, the band clearly visible in the region between \SI{200}{\metre} and \SI{1,000}{\metre} corresponds to the range of DOM-to-DOM distances between DOMs in the veto region and the reference DOMs.

Four one-dimensional PDFs are created using the information contained in the polygons indicated with dashed lines in Fig. \ref{fig:crveto_top}. Two of these PDFs are constructed from the distributions of decorrelated distance $\delta_r'$ and decorrelated time $\delta_t'$, where each is defined by a rotation $\omega = \tan^{-1}(1/c)$ in the two-dimensional space defined by $\delta_r$ and $\delta_t$. The additional PDFs are constructed from the variable $d=\sqrt{\delta_r+c\delta_t}$ and the number of pulses within each polygon. A likelihood is set up as the product of the individual ratios between the signal and background PDFs, and the likelihood ratio between these two hypotheses is used to distinguish between signal- and background-like patterns in the detector.

%Top veto
An additional top veto is applied, based on the possible coincidence between pulses in the in-ice IceCube detector and pulses recorded by the surface air-shower array, the IceTop detector~\cite{ABBASI2013188}. The veto is based on the time and lateral distance of the in-ice pulses relative to the shower axis of a moving shower plane (curvature not included) defined by the direction and timing of the event track reconstruction performed on hits in the in-ice detector only. Events with hits in IceTop that are coincident with the reconstructed track are discarded. This cut only removes 0.77\%~\cite{strom:thesis2015} of the events in the experimental data sample and has a negligible effect on the signal sample, but is nevertheless included since it removes events that are likely atmospheric muons. The relatively low efficiency of this veto can be understood in terms of the small solid angle that the IceTop detector covers.

\begin{figure}[ht!]
  \centering
  \includegraphics[width=220pt]{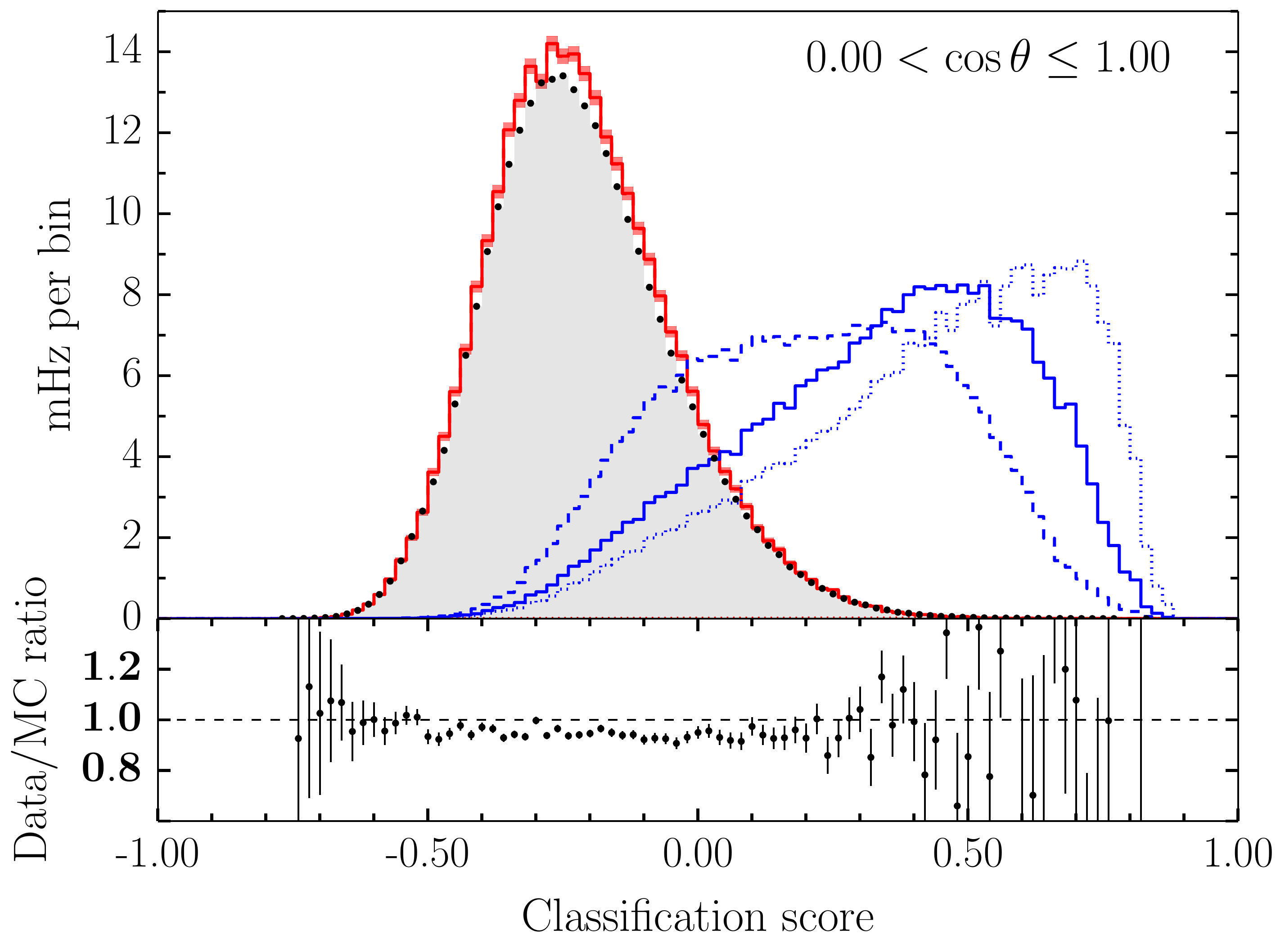}
  \caption{Classification score for the \lese selection. Experimental data is shown in black dots and illustrated by a gray shaded area. The total background simulation including both atmospheric muons $\mu$ and atmospheric muon-neutrinos $\nu_\mu$ are shown using a red solid line, with a red shaded area indicating the statistical uncertainty. Different signal hypotheses are displayed in blue lines: $\mathrm{E}_{\nu}^{-3}$ (dashed), $\mathrm{E}_{\nu}^{-2}e^{-\mathrm{E}_{\nu}/10\,\mathrm{TeV}}$ (solid), and $\mathrm{E}_{\nu}^{-2}$ (dotted), each normalized to the event rate in experimental data. The bottom panel show a comparison between experimental data and the total background simulation. Figure adapted from \cite{strom:thesis2015}.}
	\label{fig:bdt_lese}
\end{figure}

%BDT
In the next selection step a binary classifier was used, as described in Section \ref{ssec:steve}. In total, 14 features, out of 22 available features, were selected based on the feature importance metric from a preliminary classifier. Further details about these observables can be found in~\cite{strom:thesis2015}. The selected features were taken as the inputs to a binary classifier which made use of both event and observable randomization, as well as boosting and pruning.
The classifier was trained using experimental data as background while the signal sample was defined as truly down-going simulated muon neutrinos with a reconstructed interaction vertex inside the volume, defined by the dotted polygon in Fig. \ref{fig:strings_veto} and $z\leq\SI{300}{\metre}$. The resulting classification score is shown in Fig. \ref{fig:bdt_lese}.

A cut on the classification score at 0.40 is chosen to yield optimal sensitivity for an $E^{-2}$ spectrum with a \SI{10}{\tera\electronvolt} cutoff~\cite{strom:thesis2015}. As can be seen in Fig. \ref{fig:bdt_lese}, this cut removes a significant fraction of the background atmospheric muon events, enabling several advanced more time-consuming reconstructions to be performed on the remaining events, considering direction, angular uncertainty, and energy. Furthermore, shower-like events with intrinsically poor angular resolution are removed using a cut on the speed of the reconstructed particle, and a cut is made on the angular uncertainty given by the more advanced track reconstruction. As mentioned in Section \ref{ssec:reco}, this uncertainty is strongly biased towards larger values. Therefore, a bias correction is performed before removing events with uncertainties above $5^\circ$.

%% file: 4.3-Comparison.tex
\subsection{Comparison of samples} \label{ssec:comparison}
\begin{figure}
  \centering
  \includegraphics[width=220pt]{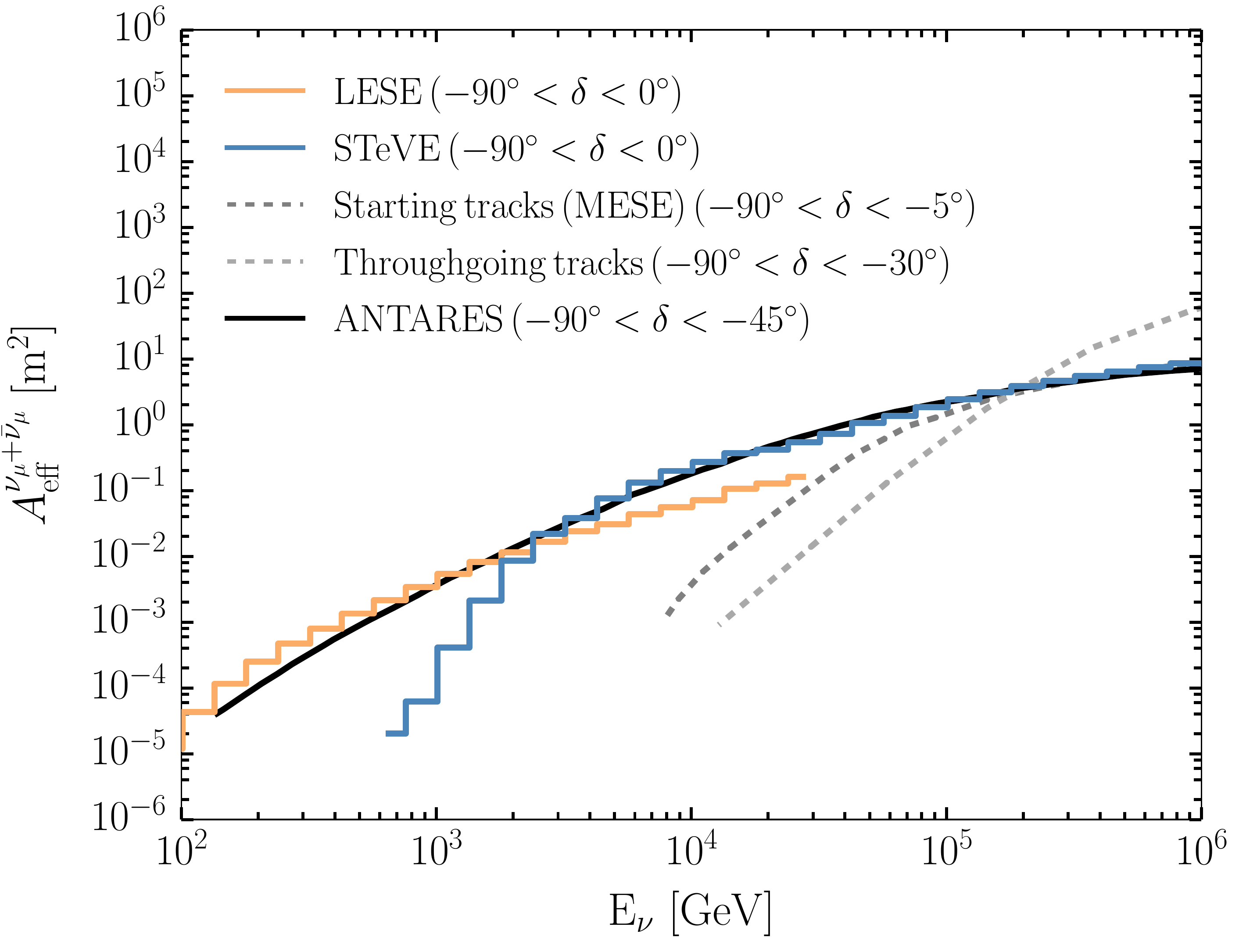}
  \caption{Effective areas of the \lese (yellow) and \steve (blue) selections compared to other IceCube selections using tracks: the throughgoing event selection~\cite{4y_PS} (dashed light gray) and the starting event selection (MESE)~\cite{MESE_paper} (dashed gray). Also shown is the effective area for ANTARES~\cite{Albert:2017ohr} (black). The effective areas are shown for a neutrino flux $\nu_{\mu}+\bar{\nu}_{\mu}$ and averaged over the solid angle in the declination range ($\delta$) indicated in the legend.}
 \label{fig:effective_area}
\end{figure}

While \steve focuses on neutrinos in the energy range of \SIrange{10}{100}{\tera\electronvolt}, \lese is optimized for neutrino energies below \SI{10}{\tera\electronvolt}. This is reflected in the effective areas, shown in Fig. \ref{fig:effective_area}. The combined effective area of \steve and \lese is comparable to the effective area of ANTARES~\cite{Albert:2017ohr} and exceeds other IceCube searches using track-like events at these energies\footnote{The effective area for ANTARES is reported in the range $-90^\circ <\delta<-45^\circ$ while the corresponding curve for LESE and STeVE is reported in the range $-90^\circ <\delta<0^\circ$.}.

\begin{figure}[b!]
  \centering
  \includegraphics[width=220pt]{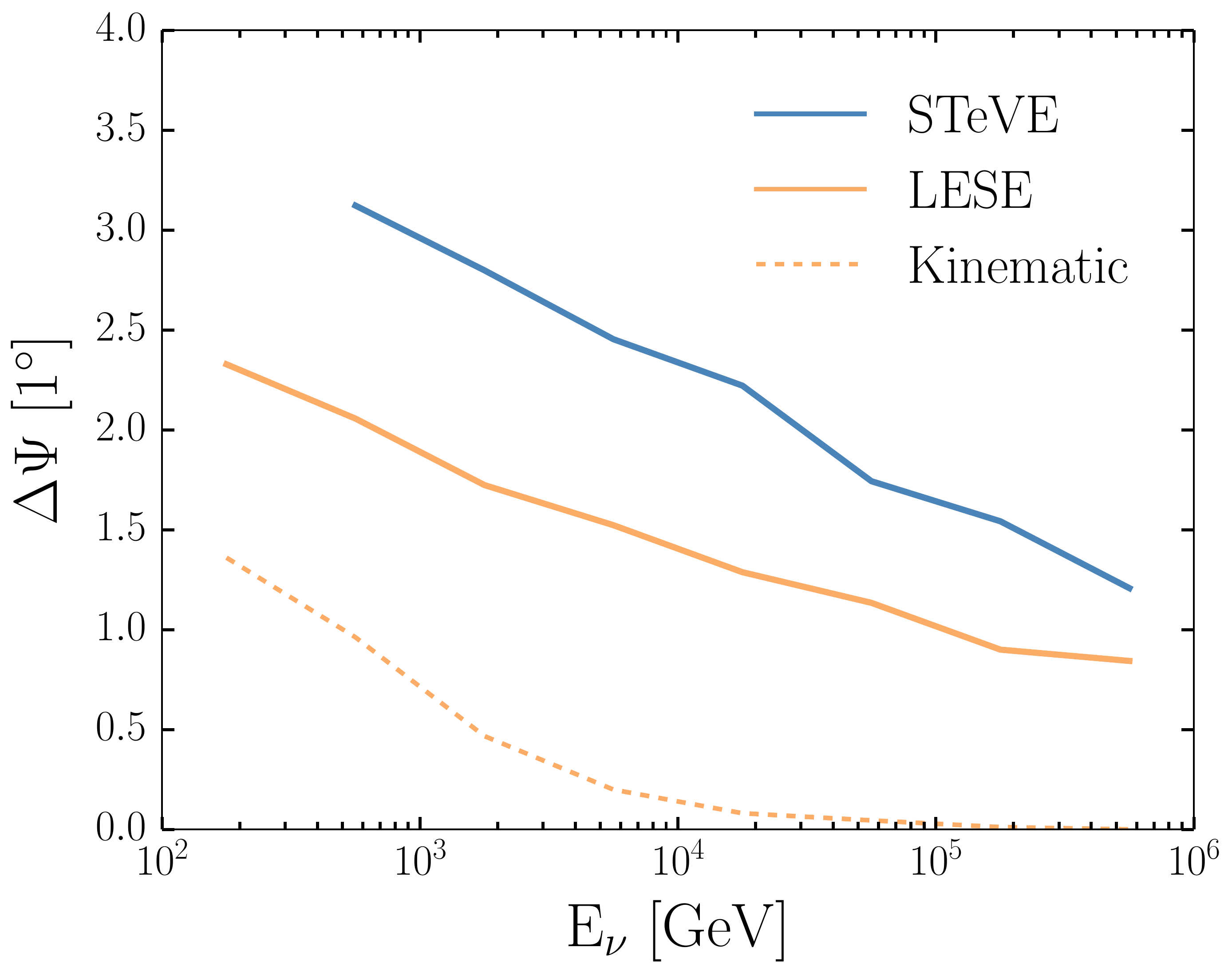}
  \caption{Median angular resolution for the \lese (yellow) and \steve (blue) selections. Also shown is the kinematic angle for the \lese selection (yellow dashed). The latter is defined as the angle between the neutrino-induced muons and the corresponding primary neutrinos and illustrates the limit to our resolution for this event selection.}
  \label{fig:angular_resolution}
\end{figure}

As discussed in Section \ref{ssec:lese}, \lese uses a large variety of cuts designed to improve the track reconstruction quality while still retaining very low-energy events. This results in a median angular resolution, defined as the median angle between the reconstructed muon and the primary neutrino, of $1.5^\circ$ for an $\mathrm{E}_\nu^{-2}$ spectrum, slightly better compared to the corresponding value for \steve. Fig.~\ref{fig:angular_resolution} shows the median angular resolution as a function of energy for \lese (yellow) and \steve (blue). Note that starting tracks, by definition, have a shorter lever arm for angular reconstruction. These events are generally not as well reconstructed as the high-energy events used in the searches with throughgoing tracks in IceCube~\cite{4y_PS, Aartsen:2016oji}.

The \steve and \lese samples are used to perform a point-source analysis, described in the following section. To ensure that there is no overlap in the targeted signal distribution, the samples are rendered mutually exclusive via a cut on the simplified energy reconstruction, described in Section~\ref{ssec:steve}, at $10^{3.7}$\,GeV. This corresponds to the crossover point in sensitivity between \lese and \steve for an $E^{-2}$ neutrino spectrum with a \SI{10}{\tera\electronvolt} cutoff. The overlap with other IceCube event samples was investigated and found to be negligible~\cite{Altmann:thesis2017}.

%% file: 5-Analysis.tex
%% ======================= APPLICATION =======================
\section{Application to a search for point-like sources} \label{sec:psanalysis}

\subsection{Experimental data and data quality} \label{ssec:data}
The IceCube detector operates in various data taking modes with a total uptime better than 99\%~\cite{IceCube:2017}. This uptime includes runs with manual overrides, for example maintenance, commissioning, and verification runs, and runs with large inactive parts of the detector. Excluding these periods, the uptime usable for physics analyses has been 97--98\% in recent years. In the case of a malfunction of a single DOM or limited parts of the detector, the data recorded is still usable in certain analyses. However, it cannot be used for event selections using vetoes since incoming muon events may ``leak in'' through the hole created, appearing to be starting inside the detector, hence mimicking the signal. In general, all runs marked \emph{good} by the detector monitoring system~\cite{IceCube:2017} were used as a baseline for the run selection. Further selection of runs was applied in each of the event selections described in Section \ref{ssec:steve} and \ref{ssec:lese}, for example by removing runs with a large number of inactive DOMs, runs shorter than 30\,min, and runs with a significant deviation in event rate compared to a sliding average.

The analysis presented in this paper uses data from the full IceCube array with 86 strings. The \steve selection uses data taken between May 2012 and May 2015, yielding 3,661 events in 1031 days of livetime. The \lese selection uses data from May 2011 to May May 2015, yielding 24,014 events in 1346 days of livetime. The corresponding event rate is 4.1$\cdot10^{-5}\,\mathrm{s}^{-1}$ (7.9$\cdot10^{-5}\,\mathrm{s}^{-1}$) for \steve and 2.1$\cdot10^{-4}\,\mathrm{s}^{-1}$ (2.3$\cdot10^{-4}\,\mathrm{s}^{-1}$) for \lese, where the value in parenthesis indicate the rate before the cut, to make the samples mutually exclusive, was applied. For consistency, a similar cut was applied to the \lese data from 2011, despite the lack of data from the \steve selection for the same time period. To avoid confirmation bias, we scrambled the event time needed to convert the azimuth angle, defined in local IceCube coordinates, to Right Ascension (R.A.) for each event until the final analysis chain was established.

\subsection{Analysis technique}
To look for clustering in the southern sky, the analysis uses an unbinned likelihood maximization similar to previous IceCube point source analyses (see for example \cite{Aartsen:2016oji} and references therein). The unbinned likelihood is constructed as the sum of the probability terms for the total number of events $N$:
\begin{equation} \label{eq:llh}
\begin{split}
\mathcal{L}(n_\mathrm{S},\gamma) = \prod_i^N \Bigg[  \frac{n_\mathrm{S}}{N}\,\mathcal{S}(|\bm{x}_\mathrm{S}-\bm{x}_i|, \mathrm{E}_i;\,\gamma) \\ + \Big( 1- \frac{n_\mathrm{S}}{N}\Big) \,\mathcal{B}(\sin\delta_i;\mathrm{E}_i) \Bigg],
\end{split}
\end{equation}
where $\bm{x}_i = (\delta_i, \alpha_i)$ denotes the reconstructed position for each event $i$ in equatorial coordinates (declination, R.A.). Furthermore, $\mathrm{E}_i$ represents the reconstructed energy and $\bm{x}_\mathrm{S}$ denotes the position of a hypothetical source S. The source is further parametrized using two parameters: the number of signal events $n_\mathrm{S}$ and the spectral index $\gamma$ with the assumed power-law spectrum $\mathrm{E}^{-\gamma}$. A formula with a cutoff is also assumed in the source list.

The spatial component of the signal hypothesis $\mathcal{S}$ is modeled using a two-dimensional Gaussian function $\mathrm{exp}(-|{\bm{x}_\mathrm{S} - \bm{x}_i |}^2 / 2\sigma_i^2)/(2\pi\sigma_i^2)$, where $\sigma_i$ represents the reconstructed angular uncertainty. The spatial component of the background hypothesis is estimated by fitting a spline function to the full experimental data sample assuming a dependence on the declination $\delta_i$ only. Furthermore, energy information is used to distinguish the soft background spectra from the typically harder signal hypotheses. Note that this addition has a limited effect for the softer signal spectra studied in this analysis. While the signal is modeled as a power-law energy spectrum, the background energy distribution is estimated from experimental data as described in \cite{Aartsen:2016oji}. We do not include a time-dependent term in the likelihood as we search for the time-averaged emission.

The total likelihood of the \steve and \lese samples combined is the product of all individual likelihoods and is maximized with respect to $n_\mathrm{S}$ and $\gamma$, yielding the best-fit values $\hat{n}_\mathrm{S}$ and $\hat{\gamma}$. Since negative $n_\mathrm{S}$ are not part of the physics scenario of neutrino sources~\cite{unbinned_llh}, we constrain $n_\mathrm{S}$ to non-negative values in the fit, i.e. $n_\mathrm{S}\geq0$. Additionally, we constrain $\gamma\in[1,6]$. The ratio of the best-fit likelihood to the likelihood under the null-hypothesis ($n_\mathrm{S} = 0$) defines the test statistic ($\mathcal{TS}$):

\begin{equation}
\mathcal{TS} = 2 \, \mathrm{log} \, \left( \frac {  \mathcal{L}  (\hat{n}_{\mathrm{S}}, \hat{\gamma}) } {  \mathcal{L} (n_{\mathrm{S}} = 0)} \right)
\label{eq:ts}
\end{equation}

The $\mathcal{TS}$ distribution follows a $\chi^2$ distribution with $n$ degrees of freedom in the limit of infinite statistics~\cite{Wilks:1938dza, wald:1943}. In this analysis $n\sim2$ since both $n_\mathrm{S}$ and $\gamma$ are allowed to float. However, the effective number of degrees of freedom is generally smaller since the parameters are partly degenerate. In addition, $\gamma$ is only defined for the case $n_\mathrm{S}>0$. 
To assess how likely it is that a certain value of $\mathcal{TS}$ is the result of a statistical fluctuation, pseudo-experiments are generated by scrambling the R.A. for the events in the experimental samples. Each scrambling results in a sky map which still accurately represents the declination dependence of signal and background events, but where any potential event clustering is washed out. The distribution of $\mathcal{TS}$ evaluated for these random skies is used to calculate the p-value of an observation being consistent with background, taking into account the fraction of overfluctuations observed~\cite{Aartsen:2016oji}.
 
\begin{figure}[ht!]
  \centering
  \includegraphics[width=220pt]{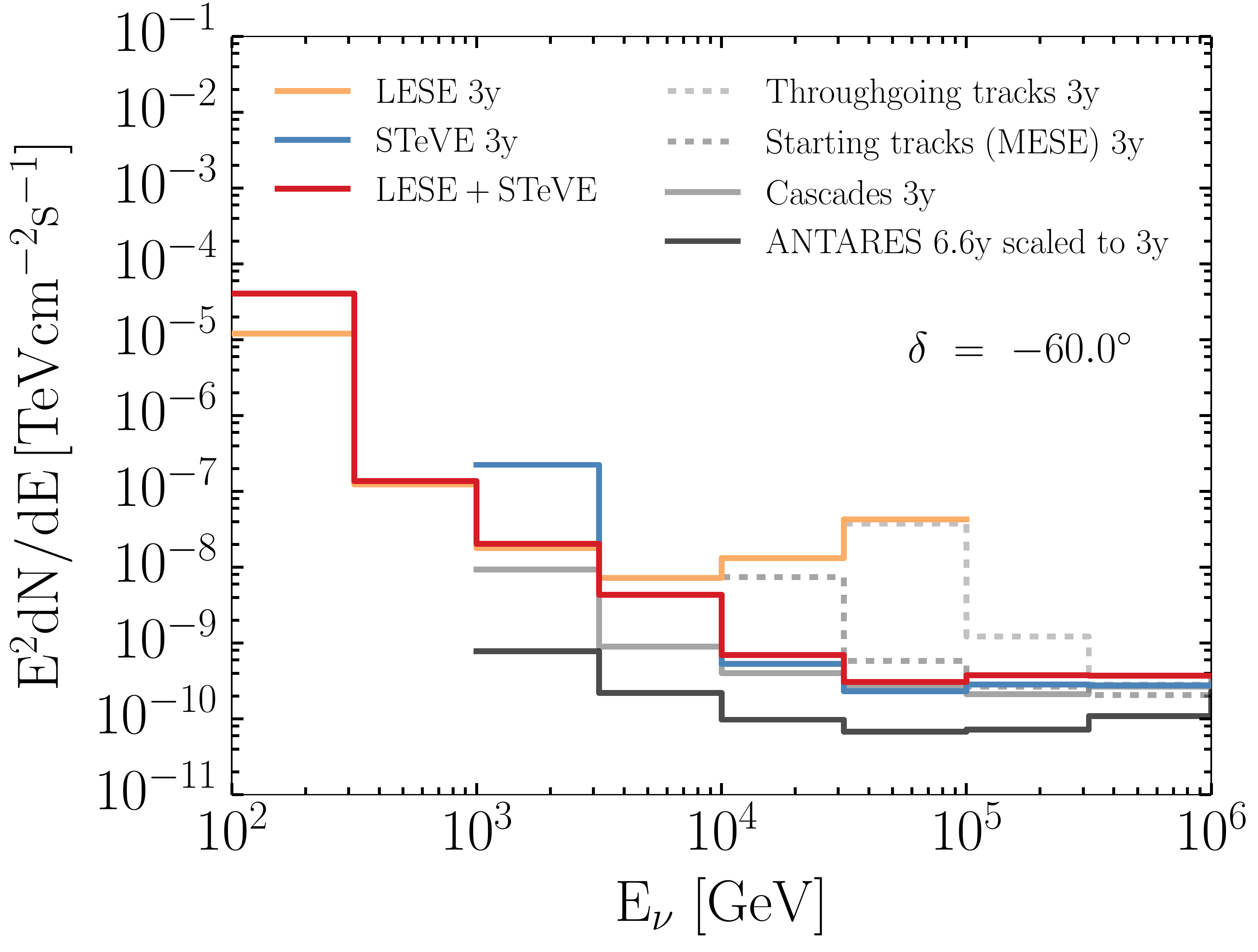}
  \caption{The differential discovery potential at \SI{-60}{\degree} declination for \lese (yellow), \steve (blue), the combined selection (\lese+\steve) (red), a cascade point-source search~\cite{mesc2017} (gray), a starting tracks search targeting higher energies (MESE)~\cite{MESE_paper} (gray dashed), throughgoing~\cite{4y_PS} (light gray dashed), all with the IceCube detector, and of the point-like source search with the ANTARES detector~\cite{antares_sens2018} based on the analysis in \cite{Albert:2017ohr} (black). In this plot, all results are calculated for an equal three year exposure.}
  \label{fig::sensitivities::differential_sensitivity}
\end{figure}

\subsection{Sensitivity and discovery potential} \label{sec:sensitivity}
The sensitivity of the analysis is defined as the flux level corresponding to a simulated source for which 90\% of pseudo-experiments with scrambled background events yield a p-value less than 0.5. Similarly, the discovery potential is defined by injecting signal events up to a flux level at which 50\% of scrambled pseudo-experiments yield a p-value corresponding to at least 5$\sigma$.
The differential discovery potentials for the individual \lese and \steve samples are shown alongside the combined sample (\lese+\steve) in Fig. \ref{fig::sensitivities::differential_sensitivity}. Note that the latter is shown after the application of the cut on the simplified energy reconstruction\footnote{Due to inaccuracies in the energy reconstruction, the separation cut in energy removes also some events far away from the cut value. This leads to a somewhat worse performance of the combined sample at low and high energies compared to the individual samples.}, as described in Section~\ref{ssec:comparison}. In addition, we show the discovery potential for a number of other IceCube searches in the southern sky as well as a recent result from ANTARES~\cite{antares_sens2018}. 
In order to compare the different event selection methods directly, we computed the differential discovery potential in the same half-decade energy bins and scaled each selection to the equivalent of three years of detector livetime. The ANTARES result for the same exposure was obtained assuming a square-root scaling of the discovery potential. Not only do \steve and \lese samples reach lower in energy than any preceding search with tracks using IceCube data in the southern sky, the samples also shows large improvement in the discovery potential up to \SI{100}{\tera\electronvolt}.

\subsection{Searches}
We perform two different searches: one unrestricted search for neutrino sources in the southern sky, not motivated by any prior information of where such a source might be located, and one search of an excess of signal-like emission at coordinates from a pre-defined list of known gamma-ray sources.

{\bf Southern sky search:} The southern sky search is performed on a HEALPix\footnote{\href{http://healpix.sourceforge.net}{http://healpix.sourceforge.net}}~\cite{healpix:2005} grid with $\sim$\SI{0.5}{\degree} spacing. The region close to the pole ($5^\circ$) is excluded due to insufficient phase space in R.A. for scrambling. The search is not motivated by prior knowledge of any sources but is limited by the angular resolution. The likelihood is evaluated at each point of the grid. 

{\bf Source list search:} \label{ssec:sourcelist} A search is performed among sources in a pre-defined list consisting of 96 astrophysical objects  in the southern hemisphere, such as supernova remnants, pulsar wind nebulae, and active galactic nuclei. This includes 84 TeVCat\footnote{\href{http://tevcat.uchicago.edu}{http://tevcat.uchicago.edu}}~\cite{Proceedings:2007aua} sources and 12 additional source candidates previously investigated by ANTARES and IceCube. The TeVCat catalog consists of published sources seen by ground-based gamma-ray experiments\footnote{We only consider sources from the catalogs ``Default Catalog'' and ``Newly Announced'', as of May 2015.}, such as VERITAS, H.E.S.S., and MAGIC. P-values are calculated for each of the 96 sources in the list.

\subsection{Results} \label{sec:results}

\begin{figure}[ht!]
  \centering
  \includegraphics[width=220pt]{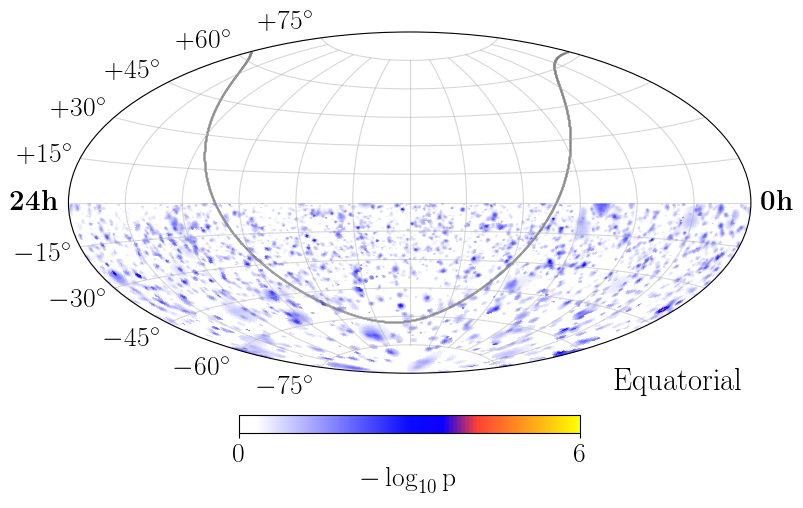}
  \caption{Pre-trial significance map in equatorial coordinates (J2000). The black line indicates the Galactic plane.}
  \label{fig:pval_skymap}
\end{figure}

\begin{figure}[ht!]
  \centering
  \includegraphics[width=220pt]{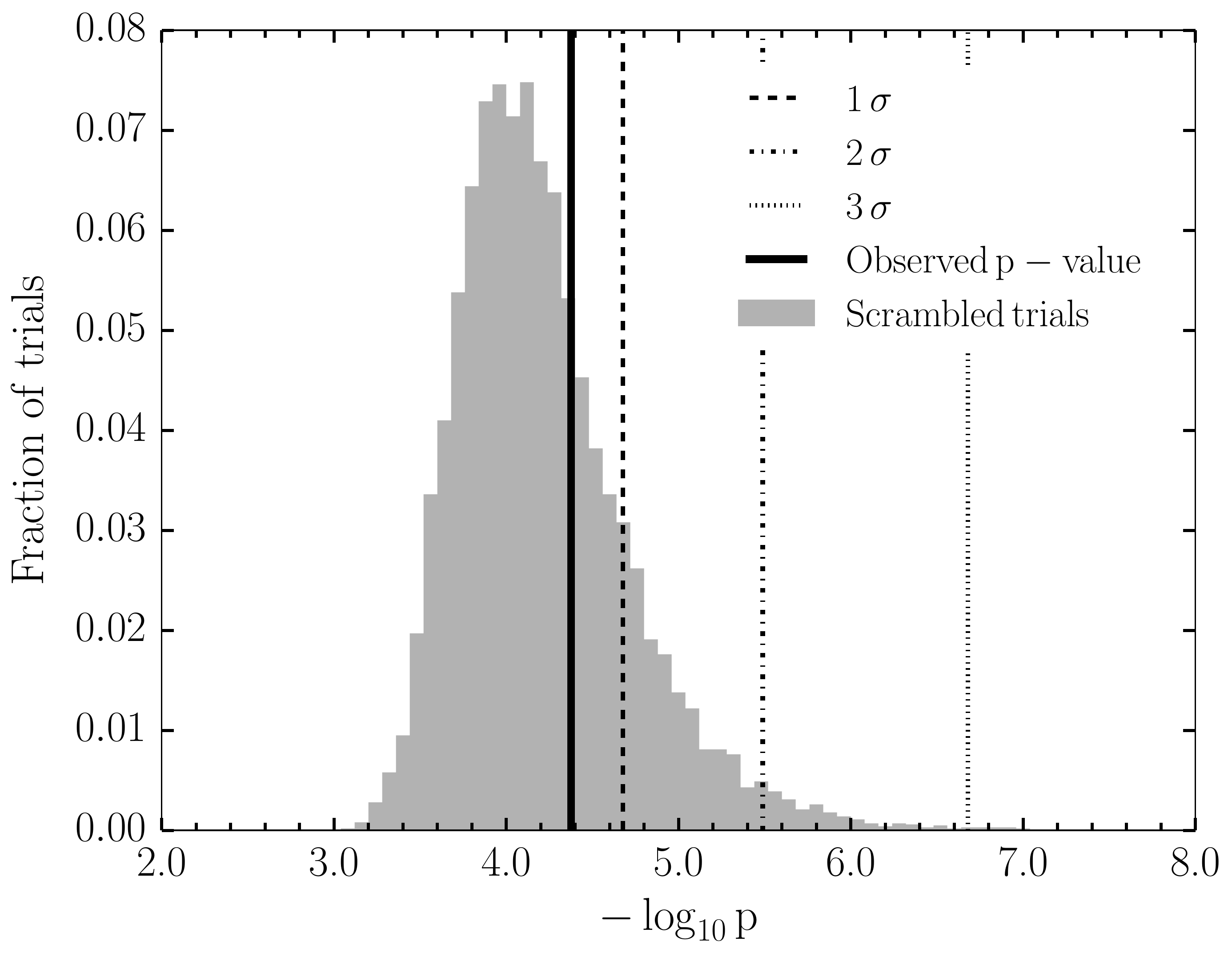}
  \caption{Distribution of the most significant p-value from the southern sky search, obtained from 10,000 randomized trials with scrambled data. The dashed vertical lines represent the 1, 2, and 3 $\sigma$ limits. The most significant p-value observed is indicated with a solid vertical line.}
  \label{fig:pval_trials}
\end{figure}

The p-values for the southern hemisphere are shown in Fig.~\ref{fig:pval_skymap}. The most significant p-value from the southern sky search, $4.2\cdot10^{-5}$, was found at $\alpha = 6.7^\circ$ and $\delta = -40.1^\circ$, with best fit parameters $\hat{n}_\mathrm{S} = 21.4$ and $\hat{\gamma} = 3.4$. Taking into account the chance of background fluctuations occurring at any position in the sky, evaluated by repeating the sky search on 10,000 randomized skies, the resulting post-trials p-value is 30.6\%. The result from the sky search is hence well compatible with the background-only hypothesis. The distribution of p-values from the random skies is shown in Fig.~\ref{fig:pval_trials}.

\begin{figure}[ht!]
  \centering
  \includegraphics[width=220pt]{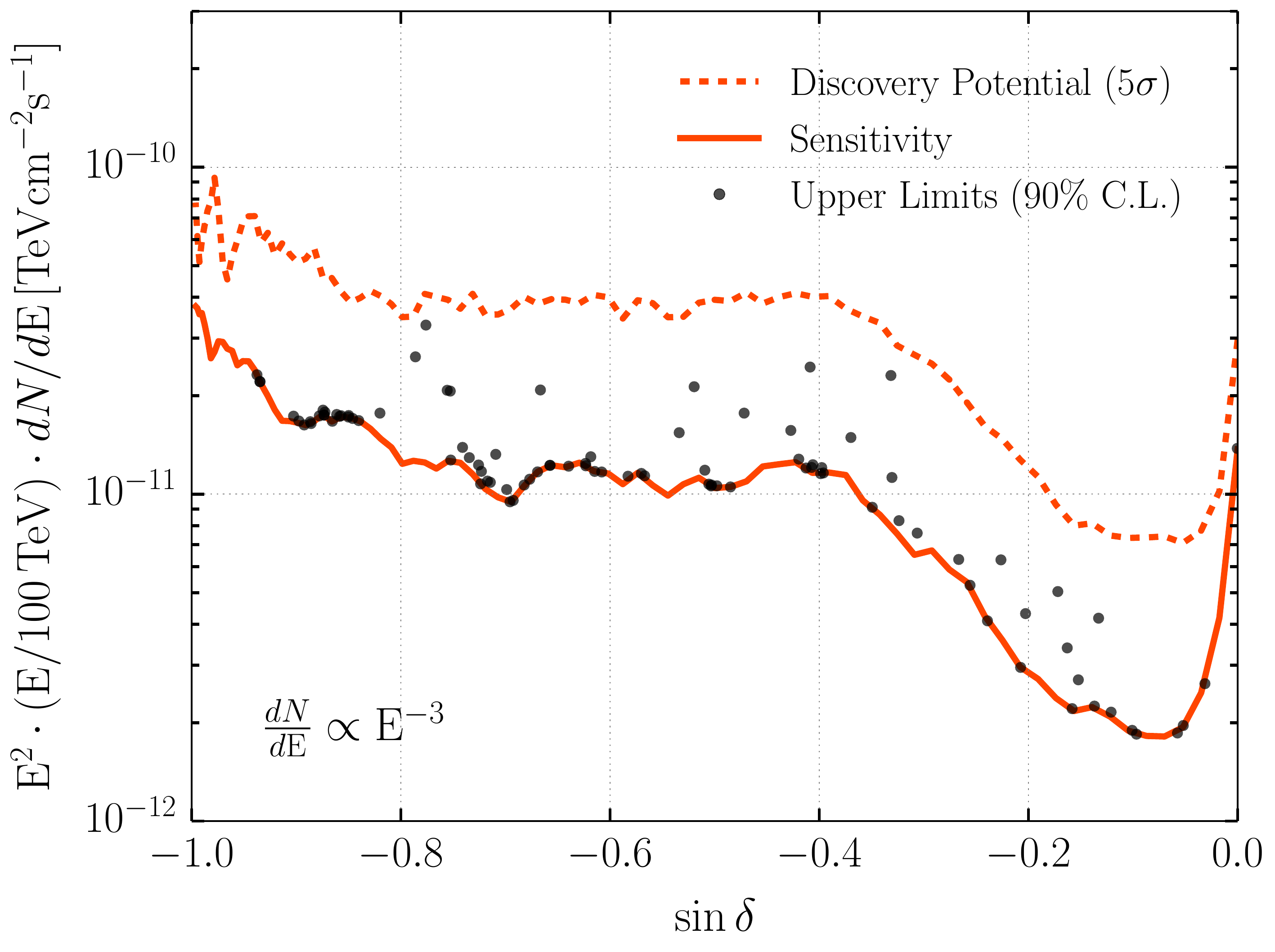}
  \caption{Sensitivity and 5$\sigma$ discovery potential as functions of declination, with flux upper limits for each object in the source catalog assuming a soft spectrum ($\gamma$ = 3).}
  \label{fig:sens_limits}
\end{figure}

The most significant source in the \emph{a priori} list was HESS J1616-508, located at $\alpha = 244.1^\circ$ and $\delta = -50.9^\circ$, with a post-trial p-value of 6.1\%. The results and upper limits at 90\% C.L., based on the frequentists approach~\cite{Neyman:1937}, for the astrophysical sources in the \emph{a priori} search list are presented in Tab.~\ref{tab:sources1} and Tab.~\ref{tab:sources2}. The upper limits derived for an $\mathrm{E}_{\nu}^{-3}$ spectra are further presented in Fig.~\ref{fig:sens_limits} along with the corresponding sensitivity and discovery potential. Since we do not consider under-fluctuations, observed values of $\mathcal{TS}$ below the median $\mathcal{TS}$ for the background-only hypothesis are reported as the corresponding median upper limit.

\subsection{Systematic uncertainties}
Since scrambled experimental data were used to estimate the statistical significance in the above analysis, the resulting p-values are not sensitive to uncertainties in the simulation of the detector or theoretical uncertainties on the atmospheric flux. However, the sensitivity and upper limits are calculated using simulated neutrino events and, hence, are affected by systematic uncertainties. The impact of systematic uncertainties is evaluated in studies where signal events from a model different from the baseline model are injected into the analysis chain.

Uncertainty in modelling the ice in the detector is one of the largest systematic uncertainties and is studied by varying two of the ice model parameters: scattering length and absorption length. Furthermore, we studied the uncertainty of the absolute efficiency of the optical modules, which describes how well the light is converted to an electrical signal in the DOM. This includes PMT quantum efficiency as well as transmittance of the optical gel and glass housing. Another uncertainty originates from the model describing the interaction of muon neutrinos with nucleons in the ice. While the baseline simulation was configured with CTEQ5~\cite{Lai:1999wy} PDFs with parton functions and cross-sections from~\cite{Gandhi:1998ri}, an alternative model tested uses the CSMS (Cooper-Sarkar, Mertsch and Sarkar) cross-section model~\cite{CooperSarkar:2011pa}. The results of varying these parameters roughly within their standard deviations leads to a total systematic uncertainty, derived as the square-root of the sum of the quadratic contributions for each of the sources of systematics studied, in the range of 15-20\% for both \steve~\cite{Altmann:thesis2017} and \lese~\cite{strom:thesis2015}.

%% file: 6-Summary.tex
\section{Summary and outlook} \label{sec:summary}

This paper presents recent advancements in data selection strategies for muon neutrinos with energies below 100 TeV from the southern sky with the IceCube detector. This includes an online filter selecting track-like events starting inside the detector as well as two new advanced veto-based strategies dubbed \lese and \steve. By using variables based on the unique event characteristics of starting tracks, both selections reduce the atmospheric background from order 100 billion triggered events to a few thousand events per year in the final event samples.

The samples were used to search for point-like neutrino sources in the southern sky at energies between 100 GeV and several TeV using four years of IceCube data. Two separate searches were performed: an unrestricted scan of the southern sky, and a search among 96 sources in a pre-defined source list. No significant deviations from the background-only hypothesis were found. After trial correction the most significant p-value from the unrestricted scan is 30.6\%. The most significant source was HESS J1616-508, with a post-trial p-value of 6.1\%, again compatible with the background-only hypothesis. Upper limits at 90\% C.L. were calculated for all the sources in the \emph{a priori} search list for a number of spectral hypotheses. 

The event selections presented in this paper improve the sensitivity and discovery potential of the IceCube detector in the southern sky for neutrinos with energies below \SI{100}{\tera\electronvolt}. In addition, they allow, for the first time, searches for point-like sources of neutrinos in the southern sky to be performed with IceCube at these energies in the track channel. While the upper limits reached for sources with assumed soft power-law spectra are of similar order of magnitude as results presented elsewhere for the cascade channel~\cite{mesc2017}, these selections, due to their considerably better pointing, would enable the localization of a sufficiently strong source. 

The samples are well suited for a large variety of analyses, including searches for extended sources and for neutrino emission in the Galactic plane.

%% file: Appendix.tex
\section{Tabulated results for sources in the \emph{a priori} search list}

This appendix contains the tabulated results for sources in the \emph{a priori} search list of astrophysical objects.

\begin{table*}[ht!]
  \begin{center}
  \resizebox{1.0\textwidth}{!}{
  \begin{tabular}{ l | c c c | c c | c c c l }
  \hline
  \hline
  \multirow{2}{*}{Source} & \multirow{2}{*}{R.A. [$^\circ$]} & \multirow{2}{*}{dec. [$^\circ$]} & \multirow{2}{*}{$-\log_{10}(\mathrm{p}$-$\mathrm{val.})$} & \multirow{2}{*}{$\hat{n}_\mathrm{S}$} & \multirow{2}{*}{$\hat{\gamma}$} & \multicolumn{3}{c}{$\Phi^{90\%}_{\nu_{\mu} +\bar{\nu}_{\mu}}\ \times\ \mathrm{TeV}\mathrm{cm^{-2}\mathrm{s}^{-1}}$} \\
&  &  &  &  &  & $\mathrm{E}_{\nu}^{-2}$ & $\mathrm{E}_{\nu}^{-2}e^{-\mathrm{E}_{\nu}/10\ \mathrm{TeV}}$ & $\mathrm{E}_{\nu}^{-3}$ \\
  \hline
IGR J18490-0000 & 282.3 & 0.0 & - & 0.0 & - & 1.11$\cdot 10^{-10}$ & 8.60$\cdot 10^{-10}$ & 1.38$\cdot 10^{-11}$\\
HESS J1848-018 & 282.1 & -1.8 & - & 0.0 & - & 3.82$\cdot 10^{-11}$ & 2.24$\cdot 10^{-10}$ & 2.63$\cdot 10^{-12}$\\
HESS J1846-029 & 281.6 & -3.0 & - & 0.0 & - & 4.37$\cdot 10^{-11}$ & 1.75$\cdot 10^{-10}$ & 1.96$\cdot 10^{-12}$\\
HESS J1843-033 & 280.8 & -3.3 & - & 0.0 & - & 4.26$\cdot 10^{-11}$ & 1.69$\cdot 10^{-10}$ & 1.86$\cdot 10^{-12}$\\
HESS J1841-055 & 280.2 & -5.5 & - & 0.0 & - & 3.08$\cdot 10^{-11}$ & 1.84$\cdot 10^{-10}$ & 1.84$\cdot 10^{-12}$\\
3C 279 & 194.0 & -5.8 & - & 0.0 & - & 3.21$\cdot 10^{-11}$ & 1.84$\cdot 10^{-10}$ & 1.89$\cdot 10^{-12}$\\
HESS J1837-069 & 279.4 & -7.0 & - & 0.0 & - & 3.84$\cdot 10^{-11}$ & 2.08$\cdot 10^{-10}$ & 2.15$\cdot 10^{-12}$\\
QSO 2022-077 & 306.4 & -7.6 & 0.11 & 7.0 & 6.0 & 6.35$\cdot 10^{-11}$ & 4.15$\cdot 10^{-10}$ & 4.17$\cdot 10^{-12}$\\
PKS 1406-076 & 212.2 & -7.9 & - & 0.0 & - & 3.99$\cdot 10^{-11}$ & 2.23$\cdot 10^{-10}$ & 2.25$\cdot 10^{-12}$\\
HESS J1834-087 & 278.7 & -8.8 & 0.37 & 1.5 & 2.5 & 4.01$\cdot 10^{-11}$ & 2.67$\cdot 10^{-10}$ & 2.70$\cdot 10^{-12}$\\
PKS 1510-089 & 228.2 & -9.1 & - & 0.0 & - & 3.39$\cdot 10^{-11}$ & 2.12$\cdot 10^{-10}$ & 2.21$\cdot 10^{-12}$\\
HESS J1832-093 & 278.2 & -9.4 & 0.26 & 2.4 & 2.3 & 5.08$\cdot 10^{-11}$ & 3.35$\cdot 10^{-10}$ & 3.39$\cdot 10^{-12}$\\
HESS J1831-098 & 277.9 & -9.9 & 0.08 & 4.5 & 2.3 & 6.46$\cdot 10^{-11}$ & 4.78$\cdot 10^{-10}$ & 5.04$\cdot 10^{-12}$\\
PKS 0727-11 & 112.6 & -11.7 & 0.18 & 7.3 & 3.6 & 6.14$\cdot 10^{-11}$ & 4.29$\cdot 10^{-10}$ & 4.31$\cdot 10^{-12}$\\
1ES 0347-121 & 57.3 & -12.0 & - & 0.0 & - & 4.28$\cdot 10^{-11}$ & 2.84$\cdot 10^{-10}$ & 2.95$\cdot 10^{-12}$\\
QSO 1730-130 & 263.3 & -13.1 & 0.12 & 6.4 & 4.6 & 5.81$\cdot 10^{-11}$ & 5.46$\cdot 10^{-10}$ & 6.30$\cdot 10^{-12}$\\
HESS J1825-137 & 276.4 & -13.8 & - & 0.0 & - & 3.72$\cdot 10^{-11}$ & 3.17$\cdot 10^{-10}$ & 4.10$\cdot 10^{-12}$\\
LS 5039 & 276.6 & -14.8 & - & 0.0 & - & 3.84$\cdot 10^{-11}$ & 4.13$\cdot 10^{-10}$ & 5.26$\cdot 10^{-12}$\\
SNR G015.4+00.1 & 274.5 & -15.5 & 0.41 & 0.5 & 3.5 & 4.53$\cdot 10^{-11}$ & 4.80$\cdot 10^{-10}$ & 6.31$\cdot 10^{-12}$\\
HESS J1813-178 & 273.4 & -17.8 & 0.44 & 0.3 & 2.6 & 4.33$\cdot 10^{-11}$ & 4.27$\cdot 10^{-10}$ & 7.60$\cdot 10^{-12}$\\
SHBL J001355.9-185406 & 3.5 & -18.9 & 0.40 & 0.9 & 3.0 & 4.72$\cdot 10^{-11}$ & 4.48$\cdot 10^{-10}$ & 8.30$\cdot 10^{-12}$\\
HESS J1809-193 & 272.6 & -19.3 & 0.22 & 3.9 & 3.8 & 5.81$\cdot 10^{-11}$ & 4.71$\cdot 10^{-10}$ & 1.12$\cdot 10^{-11}$\\
KUV 00311-1938 & 8.4 & -19.4 & 0.01 & 16.3 & 3.0 & 1.08$\cdot 10^{-10}$ & 4.75$\cdot 10^{-10}$ & 2.31$\cdot 10^{-11}$\\
HESS J1808-204 & 272.2 & -20.4 & - & 0.0 & - & 5.15$\cdot 10^{-11}$ & 5.29$\cdot 10^{-10}$ & 9.12$\cdot 10^{-12}$\\
HESS J1804-216 & 271.1 & -21.7 & 0.15 & 2.8 & 3.2 & 6.23$\cdot 10^{-11}$ & 5.94$\cdot 10^{-10}$ & 1.49$\cdot 10^{-11}$\\
W 28 & 270.4 & -23.3 & - & 0.0 & - & 4.31$\cdot 10^{-11}$ & 5.58$\cdot 10^{-10}$ & 1.16$\cdot 10^{-11}$\\
PKS 0454-234 & 74.3 & -23.4 & 0.34 & 0.1 & 2.9 & 4.33$\cdot 10^{-11}$ & 6.21$\cdot 10^{-10}$ & 1.21$\cdot 10^{-11}$\\
1ES 1101-232 & 165.9 & -23.5 & 0.37 & 0.2 & 2.8 & 4.36$\cdot 10^{-11}$ & 5.63$\cdot 10^{-10}$ & 1.16$\cdot 10^{-11}$\\
HESS J1800-240A & 270.5 & -24.0 & - & 0.0 & - & 4.55$\cdot 10^{-11}$ & 5.86$\cdot 10^{-10}$ & 1.23$\cdot 10^{-11}$\\
HESS J1800-240B & 270.1 & -24.0 & 0.37 & 0.2 & 2.9 & 4.64$\cdot 10^{-11}$ & 5.89$\cdot 10^{-10}$ & 1.21$\cdot 10^{-11}$\\
PKS 0301-243 & 45.8 & -24.1 & 0.03 & 12.8 & 3.7 & 7.76$\cdot 10^{-11}$ & 5.93$\cdot 10^{-10}$ & 2.45$\cdot 10^{-11}$\\
AP Lib & 229.4 & -24.4 & - & 0.0 & - & 5.01$\cdot 10^{-11}$ & 6.01$\cdot 10^{-10}$ & 1.20$\cdot 10^{-11}$\\
Terzan 5 & 267.0 & -24.8 & - & 0.0 & - & 5.11$\cdot 10^{-11}$ & 6.10$\cdot 10^{-10}$ & 1.28$\cdot 10^{-11}$\\
NGC 253 & 11.9 & -25.3 & 0.21 & 1.7 & 3.4 & 5.33$\cdot 10^{-11}$ & 6.12$\cdot 10^{-10}$ & 1.57$\cdot 10^{-11}$\\
SNR G000.9+00.1 & 266.8 & -28.2 & 0.13 & 2.6 & 2.8 & 4.66$\cdot 10^{-11}$ & 5.22$\cdot 10^{-10}$ & 1.77$\cdot 10^{-11}$\\
Galactic Centre & 266.4 & -29.0 & - & 0.0 & - & 3.30$\cdot 10^{-11}$ & 4.96$\cdot 10^{-10}$ & 1.05$\cdot 10^{-11}$\\
PKS 1622-297 & 246.5 & -29.9 & - & 0.0 & - & 3.41$\cdot 10^{-11}$ & 5.18$\cdot 10^{-10}$ & 1.06$\cdot 10^{-11}$\\
HESS J1741-302 & 265.2 & -30.2 & - & 0.0 & - & 3.69$\cdot 10^{-11}$ & 5.22$\cdot 10^{-10}$ & 1.07$\cdot 10^{-11}$\\
PKS 2155-304 & 329.7 & -30.2 & - & 0.0 & - & 3.71$\cdot 10^{-11}$ & 5.23$\cdot 10^{-10}$ & 1.06$\cdot 10^{-11}$\\
HESS J1745-303 & 266.3 & -30.4 & - & 0.0 & - & 3.86$\cdot 10^{-11}$ & 5.28$\cdot 10^{-10}$ & 1.07$\cdot 10^{-11}$\\
H 2356-309 & 359.8 & -30.6 & 0.39 & 0.6 & 2.7 & 4.59$\cdot 10^{-11}$ & 5.37$\cdot 10^{-10}$ & 1.18$\cdot 10^{-11}$\\
1RXS J101015.9-311909 & 152.6 & -31.3 & 0.05 & 6.4 & 3.0 & 6.67$\cdot 10^{-11}$ & 5.49$\cdot 10^{-10}$ & 2.13$\cdot 10^{-11}$\\
PKS 0548-322 & 87.7 & -32.3 & 0.15 & 5.4 & 3.3 & 5.30$\cdot 10^{-11}$ & 5.14$\cdot 10^{-10}$ & 1.54$\cdot 10^{-11}$\\
HESS J1729-345 & 262.4 & -34.5 & - & 0.0 & - & 4.35$\cdot 10^{-11}$ & 5.23$\cdot 10^{-10}$ & 1.14$\cdot 10^{-11}$\\
HESS J1731-347 & 263.0 & -34.8 & - & 0.0 & - & 4.35$\cdot 10^{-11}$ & 5.27$\cdot 10^{-10}$ & 1.16$\cdot 10^{-11}$\\
PKS 1454-354 & 224.4 & -35.6 & - & 0.0 & - & 4.06$\cdot 10^{-11}$ & 5.28$\cdot 10^{-10}$ & 1.13$\cdot 10^{-11}$\\
SNR G349.7+00.2 & 259.5 & -37.4 & - & 0.0 & - & 4.61$\cdot 10^{-11}$ & 5.44$\cdot 10^{-10}$ & 1.17$\cdot 10^{-11}$\\
PKS 0426-380 & 67.2 & -37.9 & - & 0.0 & - & 4.38$\cdot 10^{-11}$ & 5.55$\cdot 10^{-10}$ & 1.17$\cdot 10^{-11}$\\
CTB 37B & 258.5 & -38.2 & 0.32 & 0.5 & 2.5 & 4.75$\cdot 10^{-11}$ & 5.62$\cdot 10^{-10}$ & 1.30$\cdot 10^{-11}$\\
 \hline
  \hline
  \end{tabular}}
  \caption{Best-fit results and upper limits at 90\% C.L. for the astrophysical sources in the \emph{a priori} search list. The $\hat{n}_\mathrm{S}$ and $\hat{\gamma}$ columns give the best-fit values for the number of signal events and spectral index for the assumed power-law spectrum $\mathrm{E}^{-\gamma}$, respectively. The last three columns show the 90\% C.L. flux upper limits for $\nu_\mu+\bar{\nu}_\mu$, based on the classical approach~\cite{Neyman:1937}, for various source spectra. Note that the limits in the rightmost column are normalized to an $\mathrm{E}_{\nu}^{-2}$ spectrum at $\mathrm{E}=100$\,TeV.}
  \label{tab:sources1}
  \end{center}
\end{table*}

\begin{table*}[ht!]
  \begin{center}
  \resizebox{1.0\textwidth}{!}{
  \begin{tabular}{ l | c c c | c c | c c c l }
  \hline
  \hline
  \multirow{2}{*}{Source} & \multirow{2}{*}{R.A. [$^\circ$]} & \multirow{2}{*}{dec. [$^\circ$]} & \multirow{2}{*}{$-\log_{10}(\mathrm{p}$-$\mathrm{val.})$} & \multirow{2}{*}{$\hat{n}_\mathrm{S}$} & \multirow{2}{*}{$\hat{\gamma}$} & \multicolumn{3}{c}{$\Phi^{90\%}_{\nu_{\mu} +\bar{\nu}_{\mu}}\ \times\ \mathrm{TeV}\mathrm{cm^{-2}\mathrm{s}^{-1}}$} \\
&  &  &  &  &  & $\mathrm{E}_{\nu}^{-2}$ & $\mathrm{E}_{\nu}^{-2}e^{-\mathrm{E}_{\nu}/10\ \mathrm{TeV}}$ & $\mathrm{E}_{\nu}^{-3}$ \\
  \hline
HESS J1718-385 & 259.5 & -38.5 & - & 0.0 & - & 4.56$\cdot 10^{-11}$ & 5.84$\cdot 10^{-10}$ & 1.22$\cdot 10^{-11}$\\
CTB 37A & 258.6 & -38.6 & 0.37 & 0.1 & 2.5 & 4.62$\cdot 10^{-11}$ & 5.70$\cdot 10^{-10}$ & 1.24$\cdot 10^{-11}$\\
RX J1713.7-3946 & 258.4 & -39.8 & - & 0.0 & - & 3.73$\cdot 10^{-11}$ & 5.90$\cdot 10^{-10}$ & 1.22$\cdot 10^{-11}$\\
HESS J1708-410 & 257.1 & -41.1 & 0.35 & 0.1 & 2.6 & 3.76$\cdot 10^{-11}$ & 6.23$\cdot 10^{-10}$ & 1.23$\cdot 10^{-11}$\\
SN 1006-SW & 225.5 & -41.1 & - & 0.0 & - & 3.62$\cdot 10^{-11}$ & 6.01$\cdot 10^{-10}$ & 1.22$\cdot 10^{-11}$\\
SN 1006-NE & 226.0 & -41.8 & 0.04 & 5.8 & 4.0 & 6.37$\cdot 10^{-11}$ & 1.16$\cdot 10^{-9}$ & 2.08$\cdot 10^{-11}$\\
HESS J1702-420 & 255.7 & -42.0 & - & 0.0 & - & 3.53$\cdot 10^{-11}$ & 5.54$\cdot 10^{-10}$ & 1.17$\cdot 10^{-11}$\\
1ES 1312-423 & 198.7 & -42.6 & - & 0.0 & - & 3.48$\cdot 10^{-11}$ & 5.21$\cdot 10^{-10}$ & 1.11$\cdot 10^{-11}$\\
Centaurus A & 201.4 & -43.0 & - & 0.0 & - & 3.54$\cdot 10^{-11}$ & 4.95$\cdot 10^{-10}$ & 1.07$\cdot 10^{-11}$\\
PKS 0447-439 & 72.4 & -43.8 & - & 0.0 & - & 3.08$\cdot 10^{-11}$ & 4.48$\cdot 10^{-10}$ & 9.57$\cdot 10^{-12}$\\
PKS 0537-441 & 84.7 & -44.1 & - & 0.0 & - & 3.01$\cdot 10^{-11}$ & 4.44$\cdot 10^{-10}$ & 9.48$\cdot 10^{-12}$\\
HESS J1708-443 & 257.0 & -44.3 & 0.38 & 0.7 & 2.9 & 3.06$\cdot 10^{-11}$ & 4.84$\cdot 10^{-10}$ & 1.03$\cdot 10^{-11}$\\
Vela Pulsar & 128.8 & -45.2 & 0.19 & 1.9 & 3.4 & 3.96$\cdot 10^{-11}$ & 6.81$\cdot 10^{-10}$ & 1.32$\cdot 10^{-11}$\\
Vela X & 128.8 & -45.6 & 0.34 & 0.3 & 2.9 & 3.03$\cdot 10^{-11}$ & 5.46$\cdot 10^{-10}$ & 1.09$\cdot 10^{-11}$\\
Westerlund 1 & 251.7 & -45.8 & 0.33 & 0.6 & 2.9 & 3.11$\cdot 10^{-11}$ & 5.44$\cdot 10^{-10}$ & 1.10$\cdot 10^{-11}$\\
HESS J1641-463 & 250.3 & -46.3 & 0.27 & 0.7 & 3.0 & 3.66$\cdot 10^{-11}$ & 5.98$\cdot 10^{-10}$ & 1.17$\cdot 10^{-11}$\\
RX J0852.0-4622 & 133.0 & -46.4 & - & 0.0 & - & 3.12$\cdot 10^{-11}$ & 5.37$\cdot 10^{-10}$ & 1.08$\cdot 10^{-11}$\\
HESS J1640-465 & 250.2 & -46.5 & 0.28 & 0.6 & 2.9 & 3.69$\cdot 10^{-11}$ & 5.93$\cdot 10^{-10}$ & 1.23$\cdot 10^{-11}$\\
HESS J1634-472 & 248.7 & -47.3 & 0.26 & 1.6 & 3.2 & 3.56$\cdot 10^{-11}$ & 6.29$\cdot 10^{-10}$ & 1.29$\cdot 10^{-11}$\\
HESS J1632-478 & 248.0 & -47.8 & 0.24 & 1.3 & 2.7 & 3.83$\cdot 10^{-11}$ & 6.67$\cdot 10^{-10}$ & 1.39$\cdot 10^{-11}$\\
GX 339-4 & 255.7 & -48.8 & - & 0.0 & - & 3.55$\cdot 10^{-11}$ & 6.02$\cdot 10^{-10}$ & 1.27$\cdot 10^{-11}$\\
PKS 2005-489 & 302.4 & -48.8 & 0.06 & 3.0 & 2.3 & 5.00$\cdot 10^{-11}$ & 1.09$\cdot 10^{-9}$ & 2.07$\cdot 10^{-11}$\\
HESS J1626-490 & 246.5 & -49.1 & 0.06 & 4.9 & 3.6 & 5.02$\cdot 10^{-11}$ & 1.12$\cdot 10^{-9}$ & 2.08$\cdot 10^{-11}$\\
HESS J1616-508 & 244.1 & -50.9 & 0.00 & 2.7 & 1.9 & 7.85$\cdot 10^{-11}$ & 1.88$\cdot 10^{-9}$ & 3.29$\cdot 10^{-11}$\\
HESS J1614-518 & 243.6 & -51.8 & 0.01 & 1.8 & 1.8 & 5.84$\cdot 10^{-11}$ & 1.52$\cdot 10^{-9}$ & 2.63$\cdot 10^{-11}$\\
SNR G327.1-01.1 & 238.7 & -55.1 & 0.12 & 3.8 & 3.5 & 4.37$\cdot 10^{-11}$ & 9.03$\cdot 10^{-10}$ & 1.77$\cdot 10^{-11}$\\
Cir X-1 & 230.2 & -57.2 & - & 0.0 & - & 4.03$\cdot 10^{-11}$ & 8.32$\cdot 10^{-10}$ & 1.68$\cdot 10^{-11}$\\
Westerlund 2 & 155.8 & -57.8 & - & 0.0 & - & 3.98$\cdot 10^{-11}$ & 8.94$\cdot 10^{-10}$ & 1.70$\cdot 10^{-11}$\\
HESS J1026-582 & 156.7 & -58.2 & - & 0.0 & - & 3.87$\cdot 10^{-11}$ & 9.19$\cdot 10^{-10}$ & 1.74$\cdot 10^{-11}$\\
HESS J1503-582 & 225.9 & -58.2 & - & 0.0 & - & 3.86$\cdot 10^{-11}$ & 9.26$\cdot 10^{-10}$ & 1.72$\cdot 10^{-11}$\\
HESS J1018-589 & 154.4 & -59.0 & - & 0.0 & - & 3.62$\cdot 10^{-11}$ & 9.32$\cdot 10^{-10}$ & 1.74$\cdot 10^{-11}$\\
MSH 15-52 & 228.5 & -59.2 & - & 0.0 & - & 3.61$\cdot 10^{-11}$ & 9.22$\cdot 10^{-10}$ & 1.73$\cdot 10^{-11}$\\
SNR G318.2+00.1 & 224.4 & -59.5 & - & 0.0 & - & 3.56$\cdot 10^{-11}$ & 9.04$\cdot 10^{-10}$ & 1.75$\cdot 10^{-11}$\\
ESO 139-G12 & 264.4 & -59.9 & - & 0.0 & - & 3.49$\cdot 10^{-11}$ & 8.83$\cdot 10^{-10}$ & 1.67$\cdot 10^{-11}$\\
Kookaburra (PWN) & 215.0 & -60.8 & 0.29 & 0.5 & 3.0 & 3.50$\cdot 10^{-11}$ & 9.42$\cdot 10^{-10}$ & 1.78$\cdot 10^{-11}$\\
HESS J1427-608 & 217.0 & -60.9 & - & 0.0 & - & 3.36$\cdot 10^{-11}$ & 9.29$\cdot 10^{-10}$ & 1.75$\cdot 10^{-11}$\\
HESS J1458-608 & 224.5 & -60.9 & - & 0.0 & - & 3.32$\cdot 10^{-11}$ & 9.23$\cdot 10^{-10}$ & 1.75$\cdot 10^{-11}$\\
Kookaburra (Rabbit) & 214.5 & -61.0 & 0.27 & 0.6 & 3.0 & 3.43$\cdot 10^{-11}$ & 9.30$\cdot 10^{-10}$ & 1.81$\cdot 10^{-11}$\\
SNR G292.2-00.5 & 169.8 & -61.4 & - & 0.0 & - & 3.36$\cdot 10^{-11}$ & 9.46$\cdot 10^{-10}$ & 1.73$\cdot 10^{-11}$\\
HESS J1507-622 & 226.7 & -62.4 & 0.34 & 0.2 & 2.5 & 3.29$\cdot 10^{-11}$ & 8.77$\cdot 10^{-10}$ & 1.64$\cdot 10^{-11}$\\
RCW 86 & 220.7 & -62.4 & - & 0.0 & - & 3.30$\cdot 10^{-11}$ & 9.06$\cdot 10^{-10}$ & 1.67$\cdot 10^{-11}$\\
HESS J1303-631 & 195.7 & -63.2 & - & 0.0 & - & 3.15$\cdot 10^{-11}$ & 8.90$\cdot 10^{-10}$ & 1.63$\cdot 10^{-11}$\\
PSR B1259-63 & 195.7 & -63.8 & - & 0.0 & - & 3.21$\cdot 10^{-11}$ & 9.23$\cdot 10^{-10}$ & 1.67$\cdot 10^{-11}$\\
HESS J1356-645 & 209.0 & -64.5 & 0.32 & 0.4 & 2.2 & 3.27$\cdot 10^{-11}$ & 9.79$\cdot 10^{-10}$ & 1.73$\cdot 10^{-11}$\\
LHA 120 & 84.4 & -69.2 & - & 0.0 & - & 3.85$\cdot 10^{-11}$ & 1.22$\cdot 10^{-9}$ & 2.21$\cdot 10^{-11}$\\
30 Dor-C & 84.0 & -69.2 & - & 0.0 & - & 3.85$\cdot 10^{-11}$ & 1.27$\cdot 10^{-9}$ & 2.21$\cdot 10^{-11}$\\
LMC N132D & 81.3 & -69.6 & - & 0.0 & - & 3.83$\cdot 10^{-11}$ & 1.32$\cdot 10^{-9}$ & 2.32$\cdot 10^{-11}$\\
  \hline
  \hline
  \end{tabular}}
  \caption{Table continued. Best-fit results and upper limits at 90\% C.L. for astrophysical sources in the \emph{a priori} search list.}
  \label{tab:sources2}
  \end{center}
\end{table*}

%% file: ms.bbl
\begin{thebibliography}{10}
\expandafter\ifx\csname url\endcsname\relax
  \def\url#1{\texttt{#1}}\fi
\expandafter\ifx\csname urlprefix\endcsname\relax\def\urlprefix{URL }\fi
\expandafter\ifx\csname href\endcsname\relax
  \def\href#1#2{#2} \def\path#1{#1}\fi

\bibitem{Gaisser:2013bla}
T.~K. Gaisser, T.~Stanev, S.~Tilav, {Cosmic ray energy spectrum from
  measurements of air showers}, Front. Phys. China 8 (2013) 748--758 (2013).
\newblock \href {https://doi.org/10.1007/s11467-013-0319-7}
  {\path{doi:10.1007/s11467-013-0319-7}}.

\bibitem{hess_gc_pevatron}
A.~e.~a. Abramowski, {Acceleration of petaelectronvolt protons in the Galactic
  Centre}, Nature 531 (2016) 476 (2016).
\newblock \href {https://doi.org/10.1038/nature17147}
  {\path{doi:10.1038/nature17147}}.

\bibitem{Ackermann2013}
M.~Ackermann, et~al., {Detection of the characteristic pion-decay signature in
  supernova remnants}, Science 339~(6121) (2013) 807--11 (Feb. 2013).
\newblock \href {https://doi.org/10.1126/science.1231160}
  {\path{doi:10.1126/science.1231160}}.

\bibitem{Aartsen:2013jdh}
M.~G. Aartsen, et~al., {Evidence for high-energy extraterrestrial neutrinos at
  the IceCube detector}, Science 342 (2013) 1242856 (2013).
\newblock \href {https://doi.org/10.1126/science.1242856}
  {\path{doi:10.1126/science.1242856}}.

\bibitem{IceCube:2018dnn}
M.~G. Aartsen, et~al., {Multimessenger observations of a flaring blazar
  coincident with high-energy neutrino IceCube-170922A}, Science 361~(6398)
  (2018) eaat1378 (2018).
\newblock \href {http://arxiv.org/abs/1807.08816} {\path{arXiv:1807.08816}},
  \href {https://doi.org/10.1126/science.aat1378}
  {\path{doi:10.1126/science.aat1378}}.

\bibitem{IceCube:2018cha}
M.~G. Aartsen, et~al., {Neutrino emission from the direction of the blazar TXS
  0506+056 prior to the IceCube-170922A alert}, Science 361~(6398) (2018)
  147--151 (2018).
\newblock \href {http://arxiv.org/abs/1807.08794} {\path{arXiv:1807.08794}},
  \href {https://doi.org/10.1126/science.aat2890}
  {\path{doi:10.1126/science.aat2890}}.

\bibitem{Aartsen:2016oji}
M.~G. Aartsen, et~al., {All-sky search for time-integrated neutrino emission
  from astrophysical sources with 7 yr of IceCube data}, Astrophys. J. 835~(2)
  (2017) 151 (2017).
\newblock \href {https://doi.org/10.3847/1538-4357/835/2/151}
  {\path{doi:10.3847/1538-4357/835/2/151}}.

\bibitem{mesc2017}
{M.G. Aartsen et. al.}, {Search for astrophysical sources of neutrinos using
  cascade events in IceCube}, Astrophys. J. 846~(2) (2017) 136 (2017).
\newblock \href {https://doi.org/10.3847/1538-4357/aa8508}
  {\path{doi:10.3847/1538-4357/aa8508}}.

\bibitem{prl:103:221102}
R.~Abbasi, et~al., {Extending the search for neutrino point sources with
  IceCube above the horizon}, Phys. Rev. Lett. 103 (2009) 221102 (Nov 2009).
\newblock \href {https://doi.org/10.1103/PhysRevLett.103.221102}
  {\path{doi:10.1103/PhysRevLett.103.221102}}.

\bibitem{pr:d78:063004}
F.~Halzen, A.~Kappes, A.~{\'O Murchadha}, Prospects for identifying the sources
  of the {Galactic} cosmic rays with {IceCube}, Phys. Rev. D78 (2008) 063004
  (2008).
\newblock \href {https://doi.org/10.1103/PhysRevD.78.063004}
  {\path{doi:10.1103/PhysRevD.78.063004}}.

\bibitem{Vissani:2006tf}
F.~Vissani, {Neutrinos from Galactic sources of cosmic rays with known
  $\gamma$-ray spectra}, Astropart. Phys. 26 (2006) 310--313 (2006).
\newblock \href {https://doi.org/10.1016/j.astropartphys.2006.07.005}
  {\path{doi:10.1016/j.astropartphys.2006.07.005}}.

\bibitem{Kistler:2006hp}
M.~D. Kistler, J.~F. Beacom, {Guaranteed and prospective Galactic TeV neutrino
  sources}, Phys. Rev. D74 (2006) 063007 (2006).
\newblock \href {https://doi.org/10.1103/PhysRevD.74.063007}
  {\path{doi:10.1103/PhysRevD.74.063007}}.

\bibitem{MESE_paper}
{M. G. Aartsen et al.}, {Lowering IceCube's energy threshold for point source
  searches in the southern sky}, Astrophys. J. Letters 824~(2) (2016) L28
  (2016).
\newblock \href {https://doi.org/10.3847/2041-8205/824/2/L28}
  {\path{doi:10.3847/2041-8205/824/2/L28}}.

\bibitem{Altmann:thesis2017}
S.~D. Altmann, {Search for TeV neutrinos from point-like sources in the
  southern sky using four years of IceCube data}, Ph.D. thesis,
  Humboldt-Universit\"at zu Berlin (2017).
\newblock \href {https://doi.org/10.18452/17715} {\path{doi:10.18452/17715}}.

\bibitem{strom:thesis2015}
R.~Str{\"o}m, {Exploring the Universe using neutrinos: a search for point
  sources in the southern hemisphere using the IceCube Neutrino Observatory},
  Ph.D. thesis, Uppsala universitet (2015).

\bibitem{IceCube:2017}
{M.G. Aartsen et. al.}, {The IceCube Neutrino Observatory: instrumentation and
  online systems}, Journal of Instrumentation 12~(03) (2017) P03012 (2017).
\newblock \href {https://doi.org/10.1088/1748-0221/12/03/P03012}
  {\path{doi:10.1088/1748-0221/12/03/P03012}}.

\bibitem{Halzen:2010yj}
F.~Halzen, S.~Klein, {IceCube: an instrument for neutrino astronomy}, Rev. Sci.
  Instrum. 81 (2010) 081101 (2010).
\newblock \href {https://doi.org/10.1063/1.3480478}
  {\path{doi:10.1063/1.3480478}}.

\bibitem{Abbasi:2008aa}
R.~Abbasi, et~al., {The IceCube data acquisition system: signal capture,
  digitization, and timestamping}, Nucl. Instrum. Meth. A601 (2009) 294--316
  (2009).
\newblock \href {https://doi.org/10.1016/j.nima.2009.01.001}
  {\path{doi:10.1016/j.nima.2009.01.001}}.

\bibitem{Abbasi:2010vc}
R.~Abbasi, et~al., {Calibration and characterization of the IceCube
  photomultiplier tube}, Nucl. Instrum. Meth. A618 (2010) 139--152 (2010).
\newblock \href {https://doi.org/10.1016/j.nima.2010.03.102}
  {\path{doi:10.1016/j.nima.2010.03.102}}.

\bibitem{IceCube:2011ym}
R.~Abbasi, et~al., {The design and performance of IceCube DeepCore}, Astropart.
  Phys. 35 (2012) 615--624 (2012).
\newblock \href {https://doi.org/10.1016/j.astropartphys.2012.01.004}
  {\path{doi:10.1016/j.astropartphys.2012.01.004}}.

\bibitem{Aartsen:2016xlq}
M.~G. Aartsen, et~al., {Observation and Characterization of a Cosmic Muon
  Neutrino Flux from the Northern Hemisphere using six years of IceCube data},
  Astrophys. J. 833~(1) (2016) 3 (2016).
\newblock \href {http://arxiv.org/abs/1607.08006} {\path{arXiv:1607.08006}},
  \href {https://doi.org/10.3847/0004-637X/833/1/3}
  {\path{doi:10.3847/0004-637X/833/1/3}}.

\bibitem{Heckcorsika:a}
D.~Heck, J.~N. Capdevielle, G.~Schatz, T.~Thouw, {CORSIKA: A Monte Carlo Code
  to Simulate Extensive Air Showers}, Report FZKA 6019, Forschungszentrum
  Karlsruhe (1998).

\bibitem{Koehne20132070}
J.~Koehne, et~al.,
  \href{http://www.sciencedirect.com/science/article/pii/S0010465513001355}{{PROPOSAL:
  A tool for propagation of charged leptons}}, Computer Physics Communications
  184~(9) (2013) 2070 -- 2090 (2013).
\newblock \href {https://doi.org/http://dx.doi.org/10.1016/j.cpc.2013.04.001}
  {\path{doi:http://dx.doi.org/10.1016/j.cpc.2013.04.001}}.
\newline\urlprefix\url{http://www.sciencedirect.com/science/article/pii/S0010465513001355}

\bibitem{icecube_ereco}
M.~Aartsen, et~al., {Energy reconstruction methods in the IceCube Neutrino
  Telescope}, JINST 9 (2014) P03009 (2014).
\newblock \href {https://doi.org/10.1088/1748-0221/9/03/P03009}
  {\path{doi:10.1088/1748-0221/9/03/P03009}}.

\bibitem{Learned:2000sw}
J.~Learned, K.~Mannheim, {High-energy neutrino astrophysics}, Ann. Rev. Nucl.
  Part. Sci. 50 (2000) 679--749 (2000).
\newblock \href {https://doi.org/10.1146/annurev.nucl.50.1.679}
  {\path{doi:10.1146/annurev.nucl.50.1.679}}.

\bibitem{Aartsen:2013bfa}
M.~G. Aartsen, et~al., {Improvement in fast particle track reconstruction with
  robust statistics}, Nucl. Instrum. Meth. A736 (2014) 143--149 (2014).
\newblock \href {https://doi.org/10.1016/j.nima.2013.10.074}
  {\path{doi:10.1016/j.nima.2013.10.074}}.

\bibitem{reconstruction}
J.~Ahrens, et~al., {Muon Track Reconstruction and Data Selection Techniques in
  AMANDA}, Nucl. Instrum. Meth. A524 (2004) 169--194 (2004).

\bibitem{Aartsen:2013rt}
M.~G. Aartsen, et~al., {Measurement of South Pole ice transparency with the
  IceCube LED calibration system}, Nucl. Instrum. Meth. A711 (2013) 73--89
  (2013).
\newblock \href {https://doi.org/10.1016/j.nima.2013.01.054}
  {\path{doi:10.1016/j.nima.2013.01.054}}.

\bibitem{Whitehorn:2013nh}
N.~Whitehorn, J.~van Santen, S.~Lafebre, {Penalized splines for smooth
  representation of high-dimensional Monte Carlo datasets}, Comput. Phys.
  Commun. 184 (2013) 2214--2220 (2013).
\newblock \href {https://doi.org/10.1016/j.cpc.2013.04.008}
  {\path{doi:10.1016/j.cpc.2013.04.008}}.

\bibitem{Neunhoffer:2004ha}
{T. Neunh{\"o}ffer}, {Estimating the angular resolution of tracks in neutrino
  telescopes based on a likelihood analysis}, Astropart. Phys. 25 (2006) 220
  (2006).
\newblock \href {https://doi.org/10.1016/j.astropartphys.2006.01.002}
  {\path{doi:10.1016/j.astropartphys.2006.01.002}}.

\bibitem{Hulss:2010zz}
{J-P. H{\"u}l{\ss}}, {Search for neutrinos from the direction of the Galactic
  Center with the IceCube neutrino telescope}, Ph.D. thesis, Wuppertal
  University, Germany (2010).

\bibitem{Ackermann:2006pva}
M.~Ackermann, J.~Ahrens, X.~Bai, M.~Bartelt, S.~Barwick, et~al., {Optical
  properties of deep glacial ice at the South Pole}, J. Geophys. Res. 111~(D13)
  (2006) D13203 (2006).
\newblock \href {https://doi.org/10.1029/2005JD006687}
  {\path{doi:10.1029/2005JD006687}}.

\bibitem{Aartsen:2015xej}
M.~G. Aartsen, et~al., {Search for dark matter annihilation in the Galactic
  Center with IceCube-79}, Eur. Phys. J. C75~(10) (2015) 492 (2015).
\newblock \href {https://doi.org/10.1140/epjc/s10052-015-3713-1}
  {\path{doi:10.1140/epjc/s10052-015-3713-1}}.

\bibitem{Aartsen:2016pfc}
M.~G. Aartsen, et~al., {All-flavour search for neutrinos from dark matter
  annihilations in the Milky Way with IceCube/DeepCore}, Eur. Phys. J. C76~(10)
  (2016) 531 (2016).
\newblock \href {https://doi.org/10.1140/epjc/s10052-016-4375-3}
  {\path{doi:10.1140/epjc/s10052-016-4375-3}}.

\bibitem{Aartsen:2017if}
M.~G. Aartsen, et~al., {Search for annihilating dark matter in the Sun with
  3~years of IceCube data}, Eur. Phys. J. C 77~(3) (2017) 279--12 (2017).
\newblock \href {https://doi.org/10.1140/epjc/s10052-017-4689-9}
  {\path{doi:10.1140/epjc/s10052-017-4689-9}}.

\bibitem{ABBASI2013188}
{R. Abbasi et al.}, {IceTop: the surface component of IceCube}, Nuclear
  Instruments and Methods in Physics Research Section A: Accelerators,
  Spectrometers, Detectors and Associated Equipment 700~(Supplement C) (2013)
  188 -- 220 (2013).
\newblock \href {https://doi.org/https://doi.org/10.1016/j.nima.2012.10.067}
  {\path{doi:https://doi.org/10.1016/j.nima.2012.10.067}}.

\bibitem{Albert:2017ohr}
A.~Albert, et~al., {First all-flavor neutrino pointlike source search with the
  ANTARES neutrino telescope}, Phys. Rev. D96~(8) (2017) 082001 (2017).

\bibitem{4y_PS}
M.~G. Aartsen, et~al., {Searches for extended and point-like neutrino sources
  with four years of IceCube data}, Astrophys. J. 796~(2) (2014) 109 (2014).
\newblock \href {https://doi.org/10.1088/0004-637X/796/2/109}
  {\path{doi:10.1088/0004-637X/796/2/109}}.

\bibitem{unbinned_llh}
J.~Braun, J.~Dumm, F.~De~Palma, C.~Finley, A.~Karle, et~al., {Methods for point
  source analysis in high energy neutrino telescopes}, Astropart. Phys. 29
  (2008) 299--305 (2008).
\newblock \href {https://doi.org/10.1016/j.astropartphys.2008.02.007}
  {\path{doi:10.1016/j.astropartphys.2008.02.007}}.

\bibitem{Wilks:1938dza}
S.~S. Wilks, {The large-sample distribution of the likelihood ratio for testing
  composite hypotheses}, Annals Math. Statist. 9~(1) (1938) 60--62 (1938).
\newblock \href {https://doi.org/10.1214/aoms/1177732360}
  {\path{doi:10.1214/aoms/1177732360}}.

\bibitem{wald:1943}
A.~Wald, {Tests of statistical hypotheses concerning several parameters when
  the number of observations is large}, Trans. Am. Math. Soc. 54 (1943)
  426--482 (1943).
\newblock \href {https://doi.org/10.2307/1990256} {\path{doi:10.2307/1990256}}.

\bibitem{antares_sens2018}
I.~Taboada, {A view of the Universe with the IceCube and ANTARES neutrino
  telescopes}, XXVIII International Conference on Neutrino Physics and
  Astrophysics (Neutrino 2018), Heidelberg (Session Neutrino Astronomy, Part
  1), 2018 (2018).
\newblock \href {https://doi.org/10.5281/zenodo.1286919}
  {\path{doi:10.5281/zenodo.1286919}}.

\bibitem{healpix:2005}
K.~M. G{\'o}rski, et~al., {HEALPix: A Framework for High-Resolution
  Discretization and Fast Analysis of Data Distributed on the Sphere},
  Astrophys. J. 622~(2) (2005) 759 (2005).
\newblock \href {https://doi.org/10.1086/427976} {\path{doi:10.1086/427976}}.

\bibitem{Proceedings:2007aua}
S.~Wakely, D.~Horan, {TeVCat: An online catalog for Very High Energy Gamma-Ray
  Astronomy}, in: {Proceedings, 30th International Cosmic Ray Conference (ICRC
  2007)}, Vol.~3, 2007, p. 1341 (2007).

\bibitem{Neyman:1937}
J.~Neyman, {Outline of a theory of statistical estimation based on the
  classical theory of probability}, Royal Society of London Philosophical
  Transactions Series A 236 (1937) 333 (1937).
\newblock \href {https://doi.org/10.1098/rsta.1937.0005}
  {\path{doi:10.1098/rsta.1937.0005}}.

\bibitem{Lai:1999wy}
H.~Lai, et~al., {Global QCD analysis of parton structure of the nucleon: CTEQ5
  parton distributions}, Eur.Phys.J. C12 (2000) 375--392 (2000).
\newblock \href {https://doi.org/10.1007/s100529900196}
  {\path{doi:10.1007/s100529900196}}.

\bibitem{Gandhi:1998ri}
R.~Gandhi, C.~Quigg, M.~H. Reno, I.~Sarcevic, {Neutrino interactions at
  ultrahigh-energies}, Phys. Rev. D58 (1998) 093009 (1998).
\newblock \href {https://doi.org/10.1103/PhysRevD.58.093009}
  {\path{doi:10.1103/PhysRevD.58.093009}}.

\bibitem{CooperSarkar:2011pa}
A.~Cooper-Sarkar, P.~Mertsch, S.~Sarkar, {The high energy neutrino
  cross-section in the Standard Model and its uncertainty}, JHEP 08 (2011) 042
  (2011).
\newblock \href {https://doi.org/10.1007/JHEP08(2011)042}
  {\path{doi:10.1007/JHEP08(2011)042}}.

\end{thebibliography}
